\documentclass[review]{elsarticle}

\usepackage{amsmath}
\usepackage{array}
\usepackage{afterpage}
\usepackage{titlesec}
\usepackage{enumitem}
\usepackage{multirow}
\usepackage{algorithm}
\usepackage{algorithmicx}
\usepackage{algpseudocode}
\usepackage{hyperref}
\makeatletter
\def\BState{\State\hskip-\ALG@thistlm}
\makeatother

\journal{Neural Networks}

\begin{document}

\begin{frontmatter}

\title{T-Net: Encoder-Decoder in Encoder-Decoder architecture for the main vessel segmentation in coronary angiography}

%% Group authors per affiliation:
\author[KAIST]{Tae Joon Jun}
\author[AMC1]{Jihoon Kweon}
\author[AMC1]{Young-Hak Kim}
\author[KAIST]{Daeyoung Kim\corref{CORRESPONDING}}
\cortext[CORRESPONDING]{Corresponding author}
\ead{kimd@kaist.ac.kr}

%% or include affiliations in footnotes:
\address[KAIST]{
	School of Computing,
    Korea Advanced Institute of Science and Technology,\\
    34141 Daejeon, Republic of Korea}
    
\address[AMC1]{
	Division of Cardiology,
    University of Ulsan College of Medicine, Asan Medical Center,\\
    05505 Seoul, Republic of Korea}

\begin{abstract}
In this paper, we proposed T-Net containing a small encoder-decoder inside the encoder-decoder structure (EDiED). T-Net overcomes the limitation that U-Net can only have a single set of the concatenate layer between encoder and decoder block. To be more precise, the U-Net symmetrically forms the concatenate layers, so the low-level feature of the encoder is connected to the latter part of the decoder, and the high-level feature is connected to the beginning of the decoder. T-Net arranges the pooling and up-sampling appropriately during the encoder process, and likewise during the decoding process so that feature-maps of various sizes are obtained in a single block. As a result, all features from the low-level to the high-level extracted from the encoder are delivered from the beginning of the decoder to predict a more accurate mask. We evaluated T-Net for the problem of segmenting three main vessels in coronary angiography images. The experiment consisted of a comparison of U-Net and T-Nets under the same conditions, and an optimized T-Net for the main vessel segmentation. As a result, T-Net recorded a Dice Similarity Coefficient score (\textit{DSC}) of 0.815, 0.095 higher than that of U-Net, and the optimized T-Net recorded a \textit{DSC} of 0.890 which was 0.170 higher than that of U-Net. In addition, we visualized the weight activation of the convolutional layer of T-Net and U-Net to show that T-Net actually predicts the mask from earlier decoders. Therefore, we expect that T-Net can be effectively applied to other similar medical image segmentation problems.
\end{abstract}


\end{frontmatter}

%\linenumbers

\section{Introduction}
\label{ch3:introduction}
Semantic segmentation is a typical problem where deep learning technology is actively applied. Compared with classification, semantic segmentation has the advantage of visualizing the characteristics of an image because it can display a concrete region with classes of object. However, while labeling of classification is word or number level, labeling of semantic segmentation requires much time and effort for labeling because it needs to extract specific area from the image. Therefore, the most active area of semantic segmentation problem is medical image analysis. This is because the effect obtained by marking specific regions in the medical image is large even if time and effort are involved. Unlike general images, which have large number of objects to be segmented and their shape vary, medical images are captured with a specific purpose, so the number of classes for segmentation is relatively small and the image shape is fixed. Therefore, various methods for semantic segmentation is proposed to solve specific medical image segmentation problems \cite{ch3_surveymia}.

Currently, the most popular method for medical image semantic segmentation is the fully convolutional neural network (CNN) structure based on U-Net \cite{ch23_unet}. The U-Net consists of an encoder part extracting a feature from the original image and a decoder part restoring the feature to a mask image. However, since the size of the feature map continuously decreases during the encoder process, noise is generated while restoring the extracted features in the decoder process. Therefore, to minimize the loss of the original image, U-Net provides a concatenate layer that directly connects the encoder and decoder. However, due to the structural restriction of U-Net, there is only one set of convolution blocks matching the feature-map of the same size in the encoder and decoder, which has a limitation in generating a precise mask. More specifically, the high-level feature extracted from the latter encoder is connected to the beginning of the decoder, and the low-level feature is connected to the decoder near the prediction layer. Particularly, this limitation is fatal in the medical image problem where the number of classes of the object to be segmented is small, but the region of the mask has to be precisely segmented. An example of such a medical image problem is the segmenting main vessels from coronary angiography. In coronary angiography, the number of main vessels to be segmented is relatively small, but masks should be generated in the same form as the main vessel of the original image. In other words, the main vessel is identified among the various blood vessels of similar shape in the image, and the predicted region should be similar to the actual vessel shape in the original image. Therefore, from the low-level feature indicating the shape of vessels to the high-level feature specifying the main vessel, all levels of features should be considered in the mask restoring process.

In this paper, we propose a T-Net that allows various sizes of feature-maps between encoder and decoder, resulting in sophisticated semantic segmentation. The core concept of T-Net is encoder-decoder in encoder-decoder (EDiED) structure. Through EDiED structure, the size of the feature-map is increased by up-sampling in the encoding process, while the size of the feature-map is reduced by pooling in the decoding process. Thus, there are multiple sizes of feature maps in the same block, which allows for a more versatile combination when constructing the concatenated layers of the encoder-decoder. In other words, precise segmentation is possible from the beginning of the decoder by transmitting all levels of features extracted from every encoder block. We evaluated T-Net for the problem of segmenting three types of main vessels in coronary angiography. The three main vessels are the left anterior descending artery (LAD), the left circumflex artery (LCX), and the right coronary artery (RCA). We first compare the performance of U-Net and T-Net under the same conditions and visualize the intermediate convolutional layers to see how the actual weight varies from original to mask image. Then, we fine-tuned the optimized T-Net structure to show the best segmentation performance. As a result, T-Net showed 0.095 higher Dice Similarity Coefficient score (\textit{DSC}), 5.71\% higher sensitivity, 12.22\% higher precision than those of U-Net in the same experiment condition. The optimized T-Net showed an average of 0.890 \textit{DSC}, 88.32\% sensitivity and 90.50\% precision for the three types of main vessels segmentation from coronary angiography. Our T-Net is also expected to be effectively applied to other medical image segmentation problems that require precise segmentation.

The rest of this Chapter is structured as followed. In Section 2, we review the literature for vessel segmentation in coronary angiography and also briefly review CNN-based studies for semantic segmentation. Section 3 describes the basic structure of T-Net and shows examples of various models that can be derived from T-Net. Section 4 describes the optimal T-Net structure for the main vessel segmentation in coronary angiography. Section 5 evaluates the comparison of T-Net and U-Net and the performance of optimized T-Net. Finally, we conclude this study in Section 6 and discusses future plans.
%%%%%%%%%%%%%%%%%%%%%%%%%%%%%%%%%%%%%%%%%%%%%%%%%%%%%%%%%%%%%%%%%%%%%%%%%%%%%%%%%%%%%%%%%%%%%%%%%%%%%%%
\section{Related Work}
\label{ch3:related-work}
%%%%%%%%%%%%%%%%%%%%%%%%%%%%%%%%%%%%%%%%%%%%%%%%%%%%%%%%%%%%%%%%%%%%%%%%%%%%%%%%%%%%%%%%%%%%%%%%%%%%%%%
\subsection{Main vessel segmentation in coronary angiography}
\begin{figure}[!tbh]
	\centering
	\includegraphics[width=\textwidth]{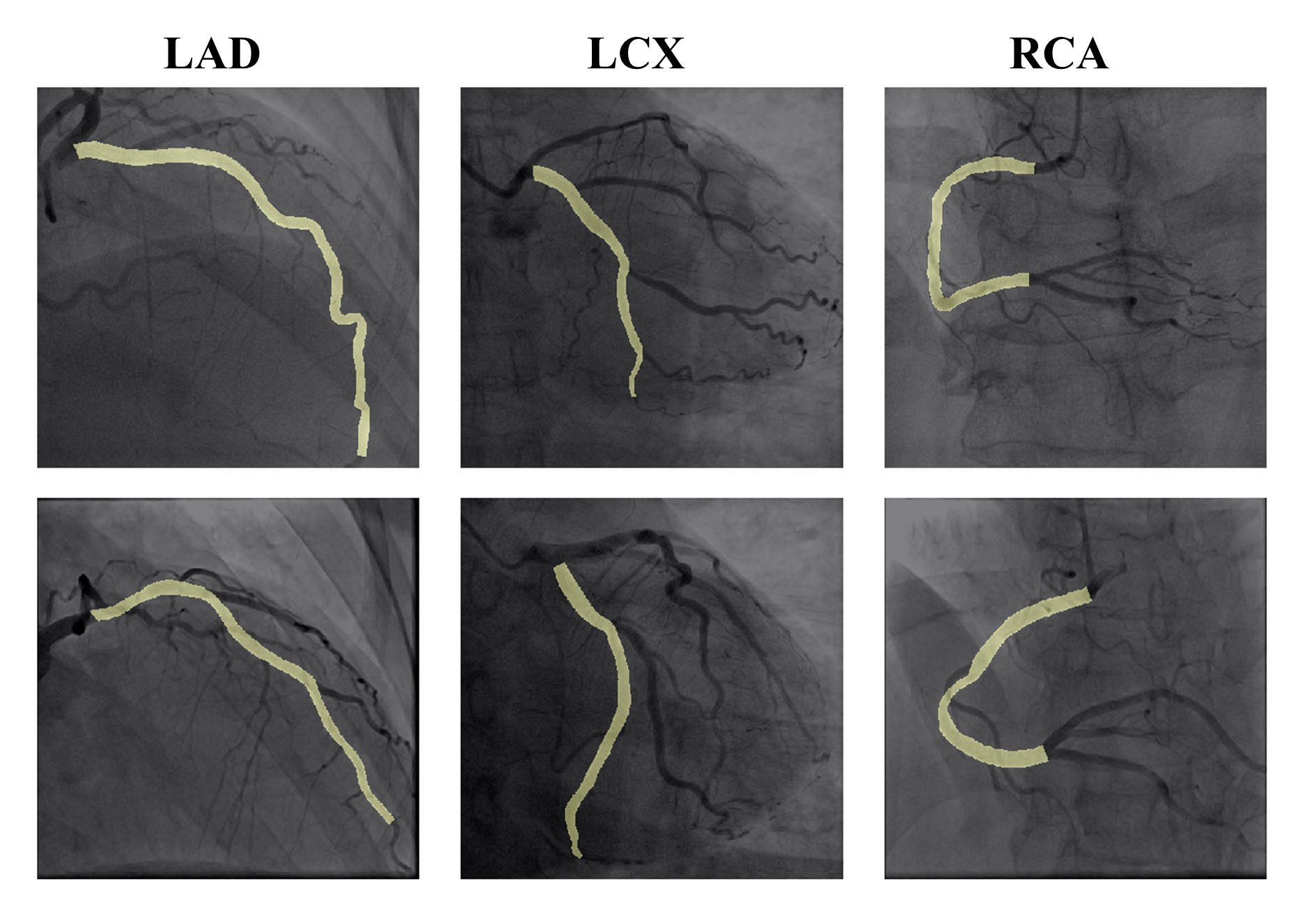}
	\caption{Three main vessels with overlapping mask images} 
	\label{img:vessel_ex}
\end{figure}

% CVD 이야기랑 main vessel segmentation이 어떤 의미를 갖는지 조사해서 설명하고
% main vessel segmentation 문제는 심장 진료에 중요한 영상인 coronary angiography에서 3 가지 type의 main vessel(LAD, LCX, RCA)를 segment하는 문제로 automated segmentation을 통해서 방사선사의 manual segmentation 작업의 burden을 줄여줄 수 있다. 
% 아산 병원에서 LAD segmentation한 것을 related work에 넣자
The World Health Organization (WHO) has announced that cardiovascular diseases (CVDs) are the leading cause of death in today's world \cite{ch3:whocvd}. More than 17 million people died of CVDs in 2016 which is an about 31\% of all deaths, and more than 75\% of these deaths occurred in low-income and middle-income countries \cite{ch3:whocvd}. Among CVDs, coronary artery disease (CAD) is the most common cause of death \cite{ch3_caddeath1}\cite{ch3_caddeath2}. In 2015, CAD affected 110 million people, resulting in 8.9 million deaths and 15.6\% of all deaths, making it the most common cause of death worldwide \cite{ch3_caddeath1}\cite{ch3_caddeath2}. The primary imaging method to observe CAD is X-ray angiography, often called coronary angiography. Especially in coronary angiography, it is important to identify the main blood vessels correctly. However, the identification of main vessels currently depends on the manual segmentation from the radiologist, requiring a lot of time and effort. Moreover, it is difficult to identify the vessel clearly because of low contrast, non-uniform illumination, and low signal to noise ratios (SNR) of X-ray angiography \cite{ch3_vessel1}. Therefore, studies on automated vessel segmentation are aimed at reducing time and cost by helping relevant experts. Among them, the main vessel segmentation is a difficult problem because it does not identify the whole blood vessels shown in the image but only the main blood vessel is segmented. Figure \ref{img:vessel_ex} shows the LAD, LCX, and RCA vessels observed in coronary angiography.

There are several studies on blood vessel segmentation in coronary angiography. Near-Esfahani proposed a CNN-based method to classify whole blood vessels in X-ray angiography \cite{ch3_vessel1}. Near-Esfahani used CNN to classify the central pixel of each patch after dividing a single image into several small patches. A total of 44 coronary angiography images of 512 x 512 size were spliced into 26 train-sets and 18 test-sets and the result was 93.5\% segmentation accuracy. Felfelian proposed a method of extracting the ROI of the coronary arteries with a Hessian filter and segmenting the blood vessel by overlapping the ROI with the flux flow measurement result \cite{ch3_vessel2}. As a result, the segmentation accuracy was about 96\% for a total of 50 x-ray angiography images. Wang proposed a method of vessel segmentation by combining Hessian matrix multi-scale filtering and region growing algorithm \cite{ch3_vessel3}. Similarly, M'hiri proposed a vessel segmentation method that combines Hessian-based vesselness information with a random walk formulation  \cite{ch3_vessel4}. Compared with existing methods such as Frangi's filter and active contour method for 9 angiography images, the AUC was 0.95. In addition, there are many other vessel segmentation approaches, but as above, they are not suitable for the main vessel segmentation problem because they segment the entire vessel in X-ray angiography \cite{ch3_vessel5}\cite{ch3_vessel6}\cite{ch3_vessel7}. In other words, a machine learning based method is needed to extract the characteristics and position of the main vessel in order to segment only the main vessel. Recently, Jo proposed the method of segmenting the LAD in coronary angiography, which is the most consistent with our study \cite{ch3_vesseljo}. Jo automatically selects the appropriate filter through the selective feature mapping (SFM) method to extract the candidate area. Then, LAD vessel segmentation is performed in the candidate area. The CNN model for segmentation is typical U-Net and is compared to U-Nets with backbone CNN using VGGNet \cite{ch23_vggnet} or DenseNet \cite{ch23_densenet}. In a total of 1,987 angiography images, 200 images were used as train-set and 1,787 images were used as test-set, and the highest result showed an average of 0.676 \textit{DSC}. Although Jo's approach utilizes U-Net and presents a novel SFM method, the segmentation performance is still low.

\subsection{CNN for semantic segmentation}
%%%%%%%%%%%%%%%%%%%%%%%%%%%%%%%%%%%%%%%%%%%%%%%%%%%%%%%%%%%%%%%%%%%%%%%%%%%%%%%%%%%%%%%%%%%%%%%%%%%%%%%
% FCN부터 U-Net, Deeplab 등 Semantic segmentation을 위한 기존의 CNN 구조를 설명하자
% 그리고 추세는 encoder뒤에 decoder가 따라오는 구조임을 말하고
% 동시에 encoder에 up-sampling이 온다거나 vice-versa 연구가 존재하지 않음을 말하자
CNN for semantic segmentation has been developed in two major directions. One is the direction of the semantic segmentation of objects in a general image. Generally, it is evaluated in PASCAL VOC \cite{ch3_pascalvoc} and MS COCO \cite{ch3_mscoco} dataset, where numbers of classes to segment are 20 and 91 respectively, excluding the background. Therefore, the feature extraction performance of the encoder is of primary importance in this direction. The first proposed CNN-based method is the fully convolutional network (FCN) proposed by Long \cite{ch3_fcn}. FCN is the model that changes the fully connected layer of the well-known classification models such as AlexNet \cite{ch23_alexnet}, VGGNet \cite{ch23_vggnet}, and GoogLeNet \cite{ch23_googlenet} to a 1x1 convolution and up-sampling the final prediction. However, there is a limitation that the process of restoring the FCN from a very small feature-map to the original mask at a single step is not accurate. Therefore, transposed convolution is proposed to overcome the limitation of FCN \cite{ch3_transposed}. Meanwhile, various studies for improving the performance in the PASCAL VOC dataset have been proposed \cite{ch3_deeplabv1} \cite{ch3_deeplabv2} \cite{ch3_pspnet} \cite{ch3_wide}. The recently proposed DeepLabv3+ \cite{ch3_deeplabv3p} takes into account the encoder-decoder architecture in previous version of DeepLabv3 \cite{ch3_deeplabv3}.

Another direction is to perform semantic segmentation in medical images. In fact, creating a mask of semantic segmentation is very costly because it is almost impossible to segment all objects that appear in a wide variety of generic images. This is why the number of classes in ImageNet \cite{ch23_imagenet}, a dataset for classification, is 1000, while PASCAL VOC and MS COCO are less than 100. However, since medical image segmentation require relatively fewer classes (tumor, vessel, organs, etc.) in typical types of images (MRI, CT, X-ray, etc.), higher performance can be obtained with a fewer number of images. In addition, because segmentation provides explainable information on medical judgment than simply classifying images, studies on medical image segmentation are very active in a wide range of medical fields \cite{ch3_surveymia} \cite{ch3_brats}. The most popular CNN based model in medical segmentation is U-Net, which is also called encoder-decoder architecture. A detailed description of U-Net follows in the next section. Also, there are variations of U-Net for medical image segmentation such as 3D U-Net \cite{ch3_3dunet}, V-Net \cite{ch3_vnet}, H-DenseUNet \cite{ch3_denseunet}, etc. However, there has been no study related to transmitting various levels of features extracted from encoders by performing multiple pooling and up-sampling in a single block to generate multi-sized feature maps.

%%%%%%%%%%%%%%%%%%%%%%%%%%%%%%%%%%%%%%%%%%%%%%%%%%%%%%%%%%%%%%%%%%%%%%%%%%%%%%%%%%%%%%%%%%%%%%%%%%%%%%%
\section{T-Net: encoder-decoder in encoder-decoder architecture}
\label{ch3:t-net}
\begin{figure}[!tbh]
	\centering
	\includegraphics[width=\textwidth]{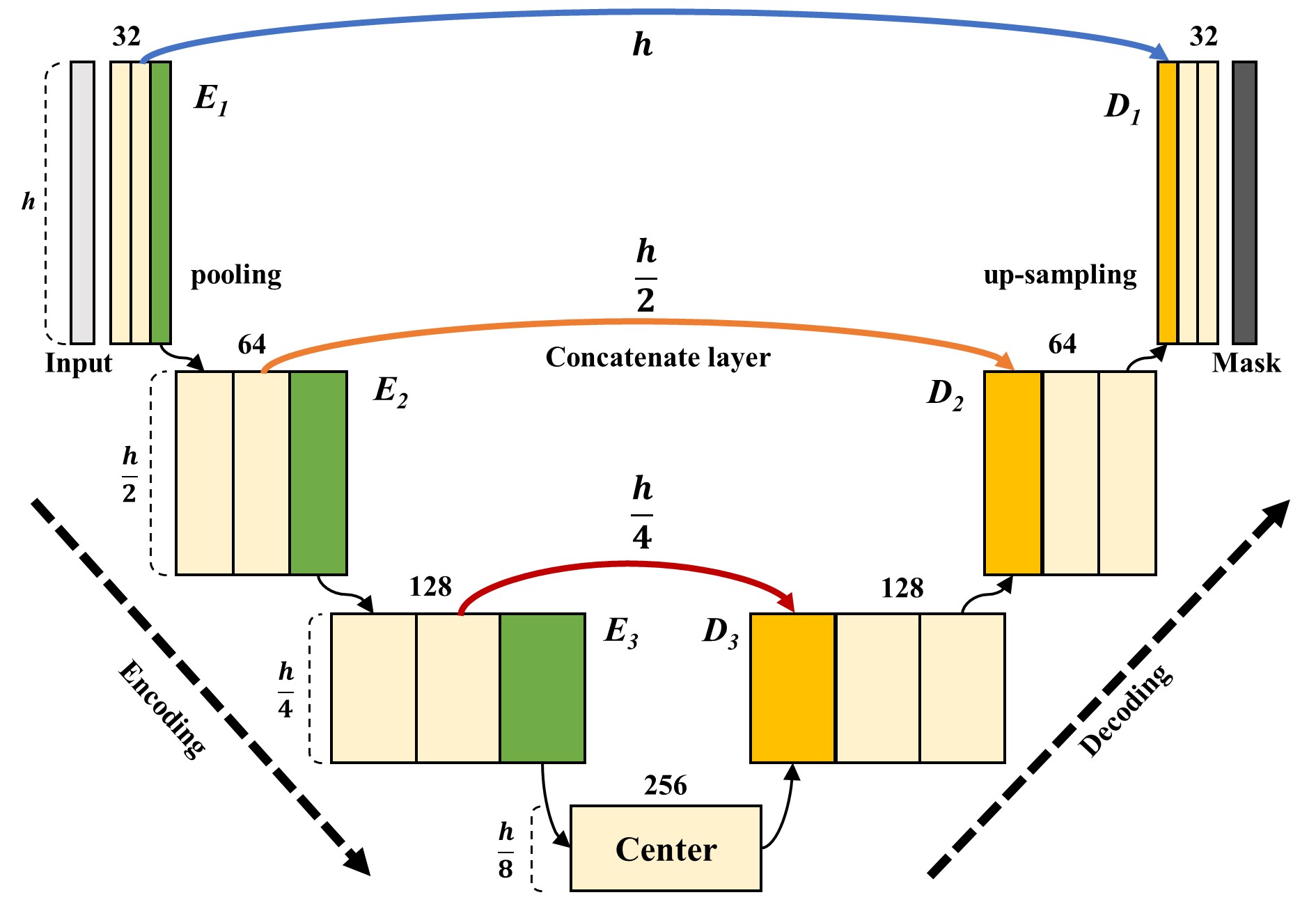}
	\caption{Basic structure of U-Net}
	\label{img:base_unet}
\end{figure}

% U-Net의 기본 구조를 먼저 설명해주면서 무엇이 문제인지 알려주자.
% T-Net의 기존 구조를 보여주자 그리고 왜 feature-map의 크기가 더 다양해지는지를 간단한 수식을 통해서 보여주자
% T-Net 구조의 다양한 variation을 설명하고 기존 모델에도 간단하게 적용할 수 있음을 보여준다
% 그리고 이러한 T-Net의 variation과 U-Net을 evaluation에서 비교할 것임을 밝힌다
Before describing T-Net in detail, we introduce the basic structure of U-Net and explain the structural limitation. U-Net is a symmetric structure, as its name implies, an encoder that extracts a feature and a decoder that restores a feature to a mask. The difference with FCN is that as the depth of the encoder increases, the number of filters increases as the general classification CNN model, and conversely, the number of filters decreases as the decoder reaches the prediction layer. Figure \ref{img:base_unet} shows the basic structure of U-Net. 

There are various deformation models of U-Net at present, but the basic structure does not deviate much from Figure \ref{img:base_unet}. That is, the encoder gradually reduces the size of the feature-map in order to extract the high-level features, while in the decoder, the size of the feature-map gradually increases to match the size of the mask. Unlike the classification in which the extracted high-level features are directly connected to the prediction layer, the segmentation requires a process of restoring to the prediction layer, which creates a noisy boundary mask different from the shape of the object in the original image. Therefore, in U-Net, there is a concatenate layer that connects feature-maps of the same size in encoder and decoder.

Suppose both the width and height of input image are \textit{h}. Let \textit{E}\textsubscript{i} be the \textit{i}-th block of the encoder, and halve the width and height of the feature-map towards \textit{E}\textsubscript{n}. Assume \textit{E}\textsubscript{1} be a direct convolutional connection to the input image to have a feature-map of size \textit{h}. Likewise, the \textit{i}-th block of the decoder is called \textit{D}\textsubscript{i}, and for convenience, let a block near the prediction layer be \textit{D}\textsubscript{1}. And \textit{D}\textsubscript{i} doubles the width and height of the feature-map by up-sampling in the direction of \textit{D}\textsubscript{1}. Therefore, \textit{E}\textsubscript{i} and \textit{D}\textsubscript{i} of U-Net have feature-map of the following sizes.
\begin{equation}\label{(equ3.1)}
	\textit{S(\textit{E}\textsubscript{i})} = \textit{S(\textit{D}\textsubscript{i})} = \frac{\textit{h}}{2\textsuperscript{i-1}}
\end{equation}
Because the concatenate layer connects the same sized convoluted layer, the concatenated layer in U-Net is only one-to-one matching of \textit{E}\textsubscript{i} and \textit{D}\textsubscript{i}. However, as the depth of the encoder becomes deeper, the high-level feature of the original image is extracted, whereas the corresponding decoder block just started restoring. On the other hand, the earliest block of the encoder extracts the low-level feature, but the matching decoder block connected is the block closest to the prediction. In other words, U-Net connects low-level features close to the prediction layer and connects high-level features far to the prediction layer. This is an inevitable limit for a single set of encoder-decoder architecture.

\subsection{Encoder-decoder in encoder-decoder (EDiED)}
\begin{figure}[!tbh]
	\centering
	\includegraphics[width=\textwidth]{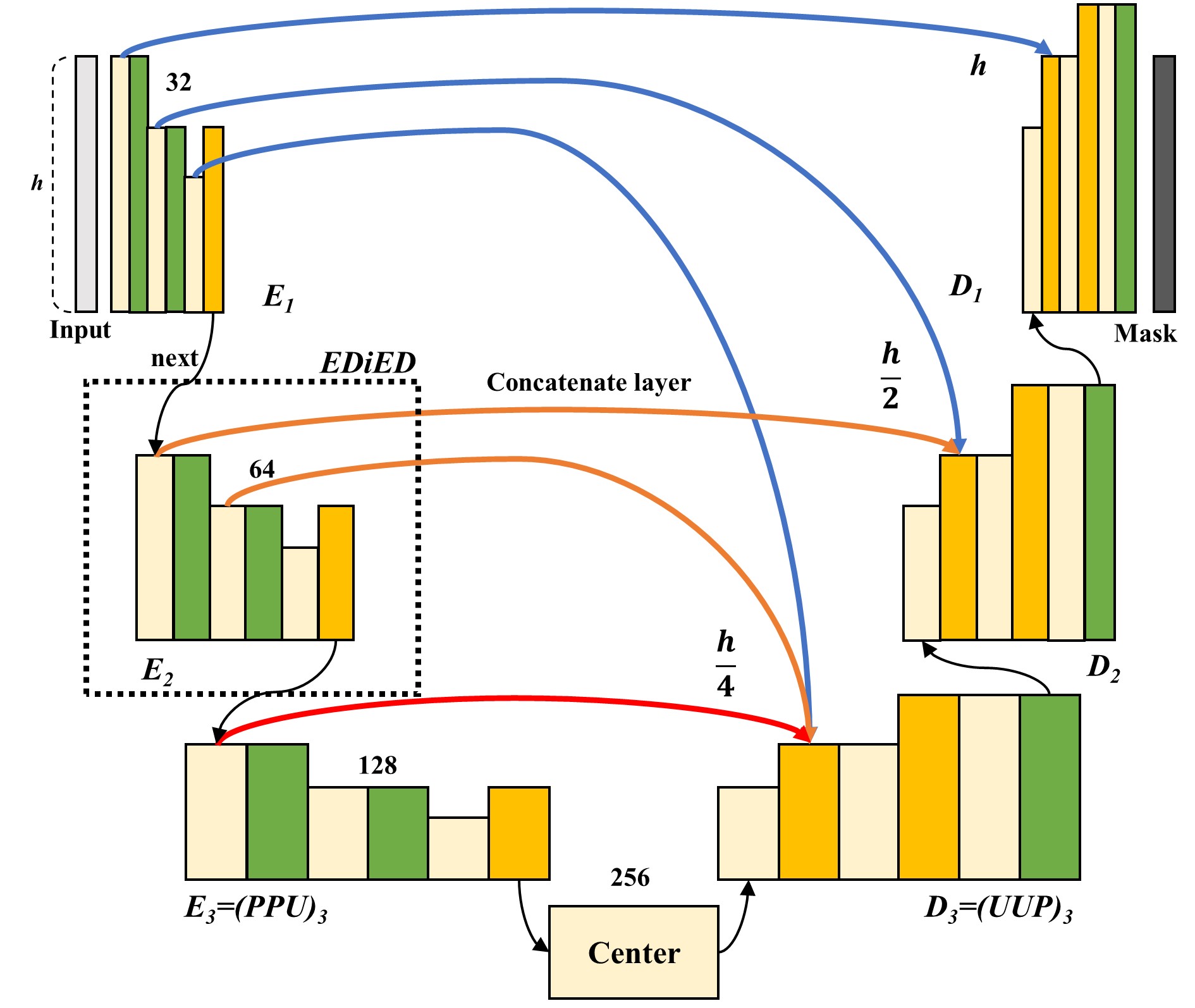}
	\caption{Structure of T3-Net with \textit{PPU} and \textit{UUP}}
	\label{img:base_t3net}
\end{figure}

In order to overcome the structural simplicity of U-Net, we propose the encoder-decoder in encoder-decoder (EDiED) architecture. The purpose of EDiED is to ensure that the various levels of features extracted from the encoder are delivered during the training of the decoder. To do this, the unmatched \textit{E}\textsubscript{i} and \textit{D}\textsubscript{j} need to be concatenated to each other. The Figure \ref{img:base_t3net} shows the simplest T-Net, which is named T3-Net.

We named it T3-Net because there are three pooling and up-sampling in the single convolution block. Likewise, if there are 5 pooling and up-sampling, it is T5-Net and if there are 5 times in encoder and 3 times in decoder, it is T53-Net. Depending on the order of pooling and up-sampling, the same T3-Net can be in various forms. We name the block in the order of \textit{P} (pooling) and \textit{U} (up-sampling) appearing in the block. That is, if \textit{E}\textsubscript{i} of T3-Net is composed of pooling, up-sampling, and pooling, it is called \textit{PUP}\textsubscript{i}, and in the case of \textit{D}\textsubscript{i}, it is called \textit{UPU}\textsubscript{i}. As can be inferred from EDiED, \textit{PUP} and \textit{UPU} can be regarded as small-scale encoder-decoders existing in encoder and decoder, respectively. Like U-Net, \textit{E}\textsubscript{i} reduces the size of the final feature-map in half, so the sum of \textit{n}(\textit{P}) and \textit{n}(\textit{U}) must be odd, and \textit{n}(\textit{P}) is one more than \textit{n}(\textit{U}). Likewise, \textit{D}\textsubscript{i} needs to double the size of the feature-map, so \textit{n}(\textit{U}) is one more than \textit{n}(\textit{P}). Therefore, \textit{E}\textsubscript{i} that can exist in T3-Net are \textit{PUP}\textsubscript{i} and \textit{PPU}\textsubscript{i}, and \textit{D}\textsubscript{i} are \textit{UPU}\textsubscript{i} and \textit{UUP}\textsubscript{i}. Assuming that \textit{P} and \textit{U} are general stride 2 pooling and up-sampling, the feature-map sizes of \textit{E}\textsubscript{i} and \textit{D}\textsubscript{i} that can exist in T3-Net are as follows.
\begin{equation}\label{(equ3.2)}
   \begin{split}
	\textit{S(\textit{E}\textsubscript{i})} = \textit{S(\textit{PPU}\textsubscript{i})} & =
	\{ \frac{\textit{h}}{2\textsuperscript{i-1}}, \frac{\textit{h}}{2\textsuperscript{i}}, \frac{\textit{h}}{2\textsuperscript{i+1}} \}\\
	\textit{S(\textit{E}\textsubscript{i})} = \textit{S(\textit{PUP}\textsubscript{i})} & =
	\{ \frac{\textit{h}}{2\textsuperscript{i-1}}, \frac{\textit{h}}{2\textsuperscript{i}} \}\\
	\textit{S(\textit{D}\textsubscript{i})} = \textit{S(\textit{UUP}\textsubscript{i})} & =
	\{ \frac{\textit{h}}{2\textsuperscript{i-3}}, \frac{\textit{h}}{2\textsuperscript{i-2}}, \frac{\textit{h}}{2\textsuperscript{i-1}} \}\\
	\textit{S(\textit{D}\textsubscript{i})} = \textit{S(\textit{UPU}\textsubscript{i})} & =
	\{ \frac{\textit{h}}{2\textsuperscript{i-2}}, \frac{\textit{h}}{2\textsuperscript{i-1}} \}\\
	\end{split}
\end{equation}

From equation \ref{(equ3.2)}, T3-Net with \textit{E}\textsubscript{i} and \textit{D}\textsubscript{i} with \textit{PPU}\textsubscript{i} and \textit{UUP}\textsubscript{i} respectively has \textit{E}\textsubscript{i-1}, \textit{E}\textsubscript{i}, \textit{E}\textsubscript{i+1} and \textit{D}\textsubscript{i+1}, \textit{D}\textsubscript{i+2} and \textit{D}\textsubscript{i+3} blocks with feature-map size \textit{h}/{2\textsuperscript{i}}. Thus, when compared to U-Net, T3-Net can add up to nine times more concatenate layers between encoder and decoder. And concatenation of these various combinations improves segmentation performance by transferring various levels of extracted features to the decoder's restore process. Figure \ref{img:base_t3net} shows the architecture of T3-Net where \textit{E}\textsubscript{i} and \textit{D}\textsubscript{i} are \textit{PPU}\textsubscript{i} and \textit{UUP}\textsubscript{i}.

\subsection{Considerations for designing T-Net}
T-Net can exist in various forms depending on the number and order of pooling and up-sampling that build up \textit{E}\textsubscript{i} and \textit{D}\textsubscript{i}. However, even if \textit{n}(\textit{P}) and \textit{n}(\textit{U}) increase, what is needed to obtain various sizes of feature-maps is continuous pooling or up-sampling. In other words, even though the depth of the block can be increased by placing \textit{P} and \textit{U} alternately, the sizes of the feature-maps generated are the same. The following equation shows the sizes of the feature-maps when alternating between \textit{P} and \textit{U}.
\begin{equation}\label{(equ3.3)}
   \begin{split}
	\textit{S(\textit{PUP}\textsubscript{i})} = 
	\textit{S(\textit{PUPUP}\textsubscript{i})} =
	\textit{S(\textit{PUPUP ... PUP}\textsubscript{i})} =
	\{ \frac{\textit{h}}{2\textsuperscript{i}}, \frac{\textit{h}}{2\textsuperscript{i-1}} \}\\
	\textit{S(\textit{UPU}\textsubscript{i})} = 
	\textit{S(\textit{UPUPU}\textsubscript{i})} =
	\textit{S(\textit{UPUPU ... UPU}\textsubscript{i})} =
	\{ \frac{\textit{h}}{2\textsuperscript{i-2}}, \frac{\textit{h}}{2\textsuperscript{i-1}} \}\\
   \end{split}
\end{equation}

On the other hand, it should also be taken into consideration that if the continuous \textit{P} or \textit{U} are long-lasting, the depth of \textit{E}\textsubscript{i} can not be deepened. For example, consider T7-Net, where \textit{E}\textsubscript{i} is \textit{PPPPUUU}\textsubscript{i}. Suppose that the size \textit{h} of the original image is 256, which we normally deal in CNN classification. In this case, the feature-map sizes of \textit{E}\textsubscript{i} are as follows.
\begin{equation}\label{(equ3.4)}
	\textit{S(\textit{PPPPUUU}\textsubscript{i})} = 
	\{ \frac{\textit{256}}{2\textsuperscript{i-1}}, \frac{\textit{256}}{2\textsuperscript{i}},  \frac{\textit{256}}{2\textsuperscript{i+1}}, \frac{\textit{256}}{2\textsuperscript{i+2}}, \frac{\textit{256}}{2\textsuperscript{i+3}} \}
\end{equation}
That is, the size of the smallest feature-map in the third encoder block becomes 4 (256/64). Empirically, it is not recommended that the size of the last feature-map of the encoder be reduced to less than 8. In the semantic segmentation problem, this is because we can extract the high-level feature as we reduce the feature-map, but it will be difficult to restore it to the mask. Also, repeated placement of pooling or up-sampling at short depths has the disadvantage of making the shape of the feature-map too simple before extracting sufficient levels of features. For example, considering the \textit{E}\textsubscript{i} of the \textit{PPPUU} structure in which the convolution and pooling layers are arranged three times, the size of the feature-maps may vary, but by reducing the size of the feature-map by one-eighth in the same block, this structure is insufficient as an encoder. Therefore, we configured only up to T5-Net in the evaluation and set the smallest feature-map size of the encoder not to be smaller than 8. And \textit{E}\textsubscript{i} in T5-Net is designed as \textit{PPUPU} structure instead of \textit{PPPUU} to avoid more than two continuous pooling layers.

\begin{figure}[!tbh]
	\centering
	\includegraphics[width=\textwidth]{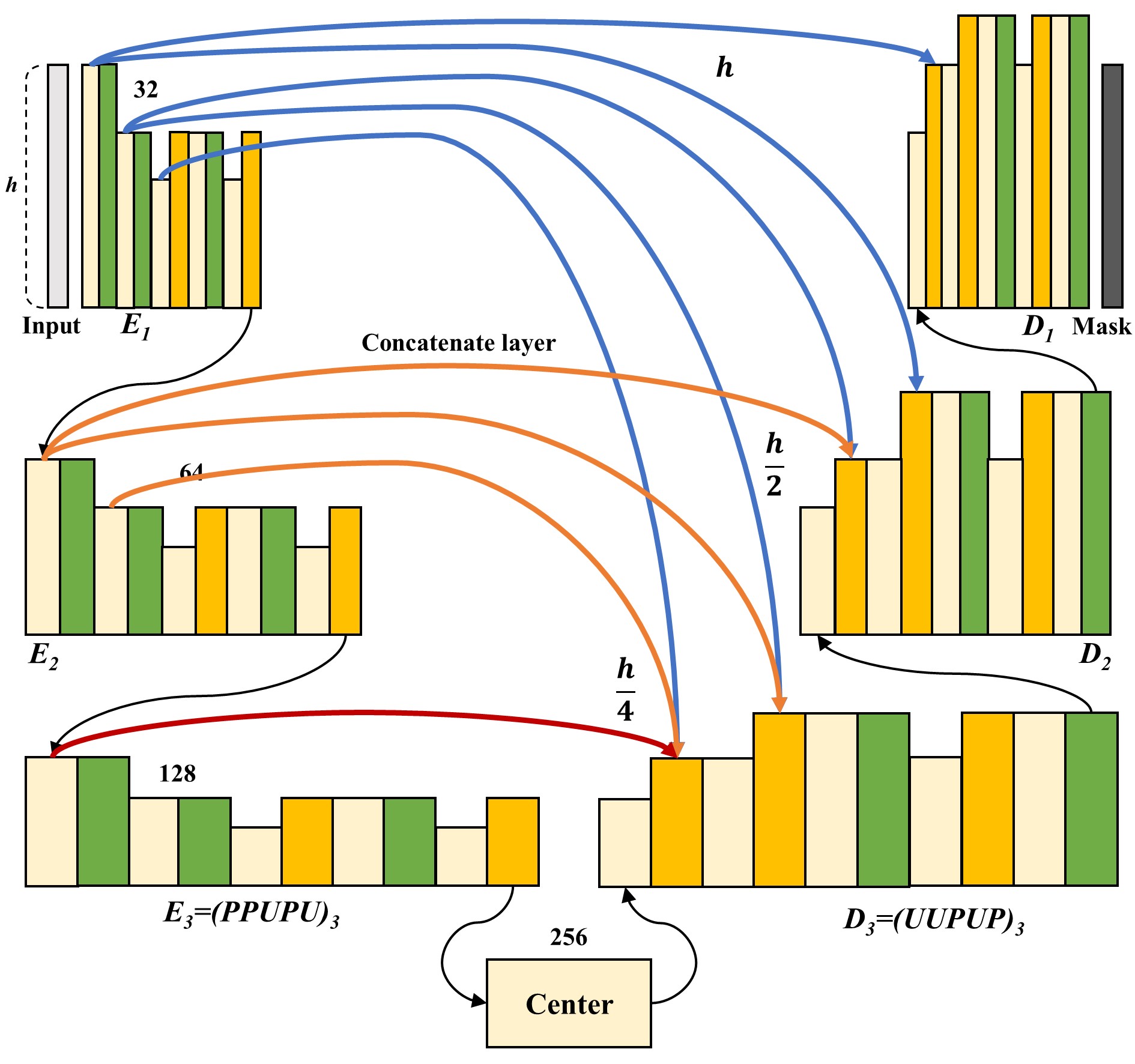}
	\caption{Structure of T5-Net with \textit{PPUPU} and \textit{UUPUP}}
	\label{img:base_t5net}
\end{figure}

The last point to consider when designing T-Net is that the \textit{P} and \textit{U} of the encoder and decoder need not be symmetric. That is, there is no problem in configuring the encoder with \textit{PPUPU} and the decoder with \textit{UUP}. However, since the purpose of EDiED is to transfer the various levels of features extracted from the encoder to the decoder, it is recommended that the length of the block constituting the encoder is longer than the decoder. In other words, T53-Net delivers more feature levels to the decoder than T35-Net. Therefore, we added T53-Net as a comparison with U-Net. In T53-Net, the encoder block is composed of \textit{PPUPU} and the decoder block is composed of \textit{UUP}. As a result, the T-Nets we have configured for comparison with U-Net are T3-Net, T5-Net, and T53-Net. The structure of T3-Net is in Figure \ref{img:base_t3net}, and the structure of T5-Net is shown in Figure \ref{img:base_t5net}. T53-Net is the same as replacing T5-Net decoder with \textit{UUP} instead of \textit{UUPUP} in Figure \ref{img:base_t5net}.
%%%%%%%%%%%%%%%%%%%%%%%%%%%%%%%%%%%%%%%%%%%%%%%%%%%%%%%%%%%%%%%%%%%%%%%%%%%%%%%%%%%%%%%%%%%%%%%%%%%%%%%
\section{Optimized T-Net for main vessel segmentation}
\label{ch3:optimized-t-net}
\begin{figure}[!tbh]
	\centering
	\includegraphics[width=\textwidth]{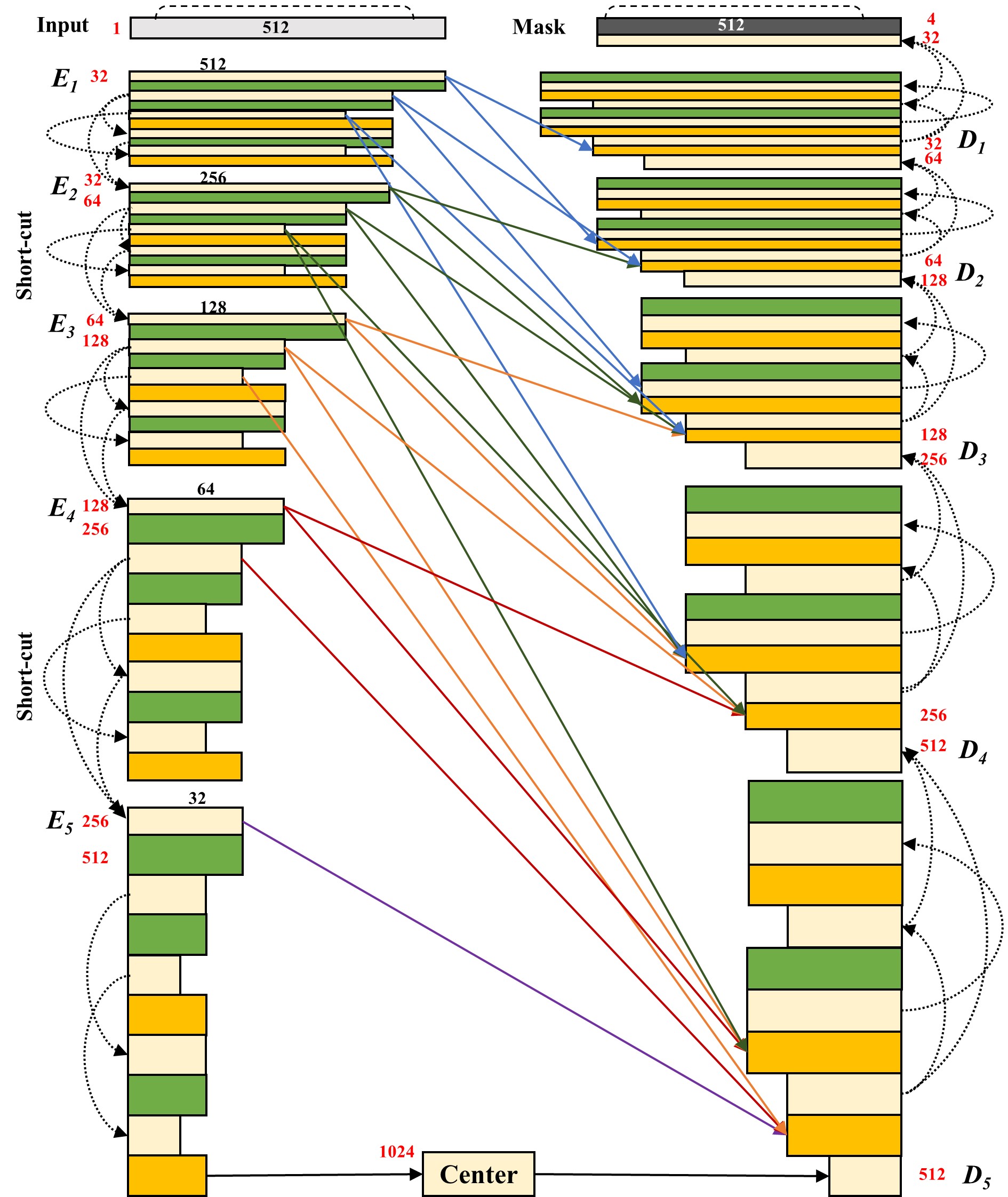}
	\caption{Optimized T-Net structure for main vessel segmentation}
	\label{img:base_optnet}
\end{figure}

% main vessel segmentation을 위한 optimization 과정을 보인다
% 여기에는 shortcut을 추가하는 방법도 설명하고, loss를 1 + 0.1 CEloss - DSC로 하는 이유도 설명하자
In this section, we describe an optimized T-Net structure for main vessel segmentation in coronary angiography. Figure \ref{img:base_optnet} shows the detailed structure of optimized T-Net. 

The overall process of glaucoma detection is as follows. First, angiography images are augmented for regularization of the model. And we train the augmented images on optimized T-Net by the mini-batch size. The model is improved to minimize the validation loss and finally, the performance of the model is evaluated using the test-set. 

\subsection{Data augmentation}
\begin{figure}[!tbh]
	\centering
	\includegraphics[width=\textwidth]{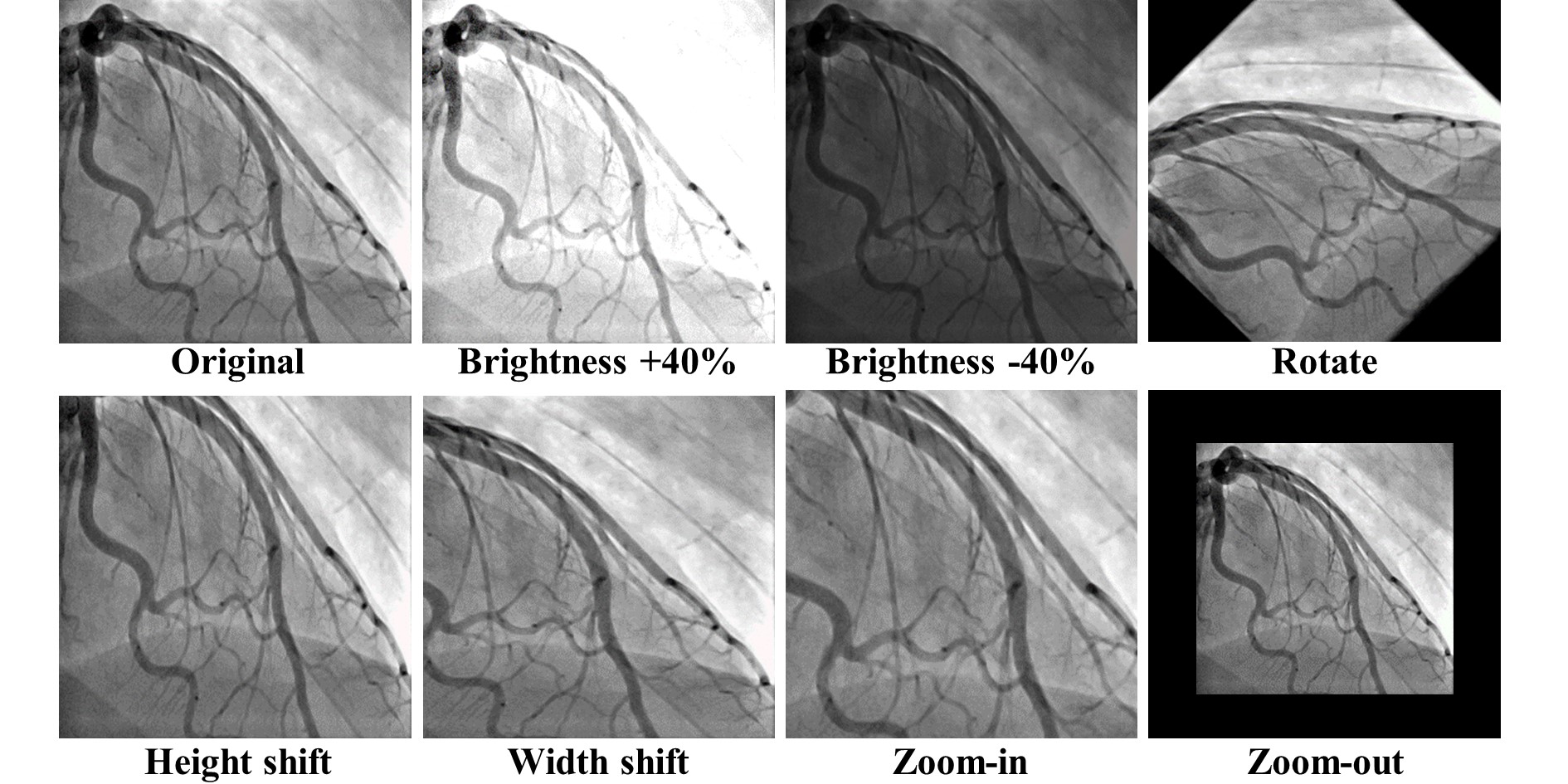}
	\caption{Vessel images result from data augmentation}
	\label{img:augment_vessel}
\end{figure}

Our data consists of 4,700 grayscale images with a size of 512 x 512. This is a small number compared to general datasets such as ImageNet \cite{ch23_imagenet} or MS COCO \cite{ch3_mscoco}, and without proper augmentation, the model will inevitably face overfitting problem. Fortunately, because coronary angiography is taken in a defined form with a specific purpose, overfitting can be avoided with augmentation even with 4,700 images. The effect of preventing overfitting can also be confirmed through the training loss of the optimized T-Net in the result section. In addition, we train the model by resizing an image in a classification problem, but since the prediction of segmentation is done by pixel levels, resizing is not recommended if GPU memory allows. Therefore, we used the size of the input image as the original 512 x 512. Our image augmentation policy is as follows. First, we zoom-in and zoom-out an image at a random ratio within $\pm$20\%. And the height and width of the image are shifted at a random ratio within $\pm$20\% of image size 512 x 512. Next, we rotate the angiography image within $\pm$30\textsuperscript{$\circ$} at random rates. Finally, because the brightness of the angiography image can vary, the brightness is also changed within $\pm$40\% at random rates. Figure \ref{img:augment_vessel} shows images when each augmentation policy is applied to a single image at a maximum rate.

\subsection{T-Net optimization for main vessel segmentation}
In this section, we describe the specific structure and fine-tuning parameters of optimized T-Net for main vessel segmentation. First, we set the overall T-Net structure to T5-Net. This is because we compared U-Net, T3-Net, T5-Net, and T53-Net under the same conditions, and as a result, T5-Net showed the best performance. However, the image used in the comparison is resized to 256 x 256 in order to shorten the training time, and the structure of optimized T-Net is advanced from that of T5-Net. The structures of T3-Net, T5-Net, and T53-Net designed for comparison with U-Net are described in detail in the result section.

The optimized T-Net consists of five encoder blocks (\textit{E}\textsubscript{i}) and decoder blocks (\textit{D}\textsubscript{i}), respectively. \textit{E}\textsubscript{i} is composed of \textit{PPUPU}\textsubscript{i}, and \textit{D}\textsubscript{i} is composed of \textit{UUPUP}\textsubscript{i}. Each pooling or up-sampling layer is preceded by a convolutional layer (\textit{C}). That is, the actual \textit{PPUPU} is \textit{CPCPCUCPCU}, but it is called \textit{PPUPU} for convenience. Let \textit{P}\textsubscript{\textit{E}\textsubscript{i}}\textsuperscript{k} be the k-th pooling or up-sampling layer that appears in \textit{E}\textsubscript{i}, and similarly define \textit{U}\textsubscript{\textit{E}\textsubscript{i}}\textsuperscript{k}, \textit{P}\textsubscript{\textit{D}\textsubscript{i}}\textsuperscript{k}, \textit{U}\textsubscript{\textit{E}\textsubscript{i}}\textsuperscript{k}, where i $\in$ \{1,2,3,4,5\} and k $\in$ \{1,2,3,4,5\}. The sizes of the feature-maps constituting \textit{E}\textsubscript{1} is as follows.
\begin{equation}\label{(equ3.4)}
	\textit{S}(\textit{E}\textsubscript{1}) = \textit{S}(\{\textit{P}\textsubscript{\textit{E}\textsubscript{1}}\textsuperscript{1}, \textit{P}\textsubscript{\textit{E}\textsubscript{1}}\textsuperscript{2}, \textit{U}\textsubscript{\textit{E}\textsubscript{1}}\textsuperscript{3}, \textit{P}\textsubscript{\textit{E}\textsubscript{1}}\textsuperscript{4}, \textit{U}\textsubscript{\textit{E}\textsubscript{1}}\textsuperscript{5}\}) =
	\{512, 256, 128, 256, 128\}
\end{equation}
Likewise, the sizes of feature-maps from \textit{E}\textsubscript{2} to \textit{E}\textsubscript{5} are as follows.
\begin{equation}\label{(equ3.5)}
    \begin{split}
	\textit{S}(\textit{E}\textsubscript{2}) & = \{256, 128, 64, 128, 64\} \\
	\textit{S}(\textit{E}\textsubscript{3}) & = \{128, 64, 32, 64, 32\} \\
	\textit{S}(\textit{E}\textsubscript{4}) & = \{64, 32, 16, 32, 16\} \\
	\textit{S}(\textit{E}\textsubscript{5}) & = \{32, 16, 8, 16, 8\} \\
	\end{split}
\end{equation}
As previously defined, the decoder block becomes D1 near the prediction, and the block close to the convolutional layers of center becomes D5. Therefore, the sizes of feature-maps from \textit{D}\textsubscript{5} to \textit{D}\textsubscript{1} are as follows.
\begin{equation}\label{(equ3.6)}
    \begin{split}
	\textit{S}(\textit{D}\textsubscript{5}) & = \{16, 32, 64, 32, 64\} \\
	\textit{S}(\textit{D}\textsubscript{4}) & = \{32, 64, 128, 64, 128\} \\
	\textit{S}(\textit{D}\textsubscript{3}) & = \{64, 128, 256, 128, 256\} \\
	\textit{S}(\textit{D}\textsubscript{2}) & = \{128, 256, 512, 256, 512\} \\
	\textit{S}(\textit{D}\textsubscript{1}) & = \{256, 512, 1024, 512, 1024\} \\
	\end{split}
\end{equation}
Since the \textit{S}(\textit{P}\textsubscript{\textit{D}\textsubscript{1}}\textsuperscript{5}) is 1024, the size of next convolutional layer's feature-maps become 512. Therefore, connecting this convolutional layer to the final convolutional layer with four filters and applying a softmax function to the result, a feature-map of size 512 x 512 x 4 is generated. This represents the background, LAD, LCX, and RCA scores for the 512 x 512 size mask. And each of these scores is compared with the actual class as pixel-by-pixel, and the total loss is used for the gradient calculation of the next epoch. Next, we explain how the concatenate layers are constructed based on the above feature-maps sizes and describe the specific parameters for fine-tuning.

\subsubsection{Concatenate layers in optimized T-Net}
\begin{figure}[!tbh]
	\centering
	\includegraphics[width=\textwidth]{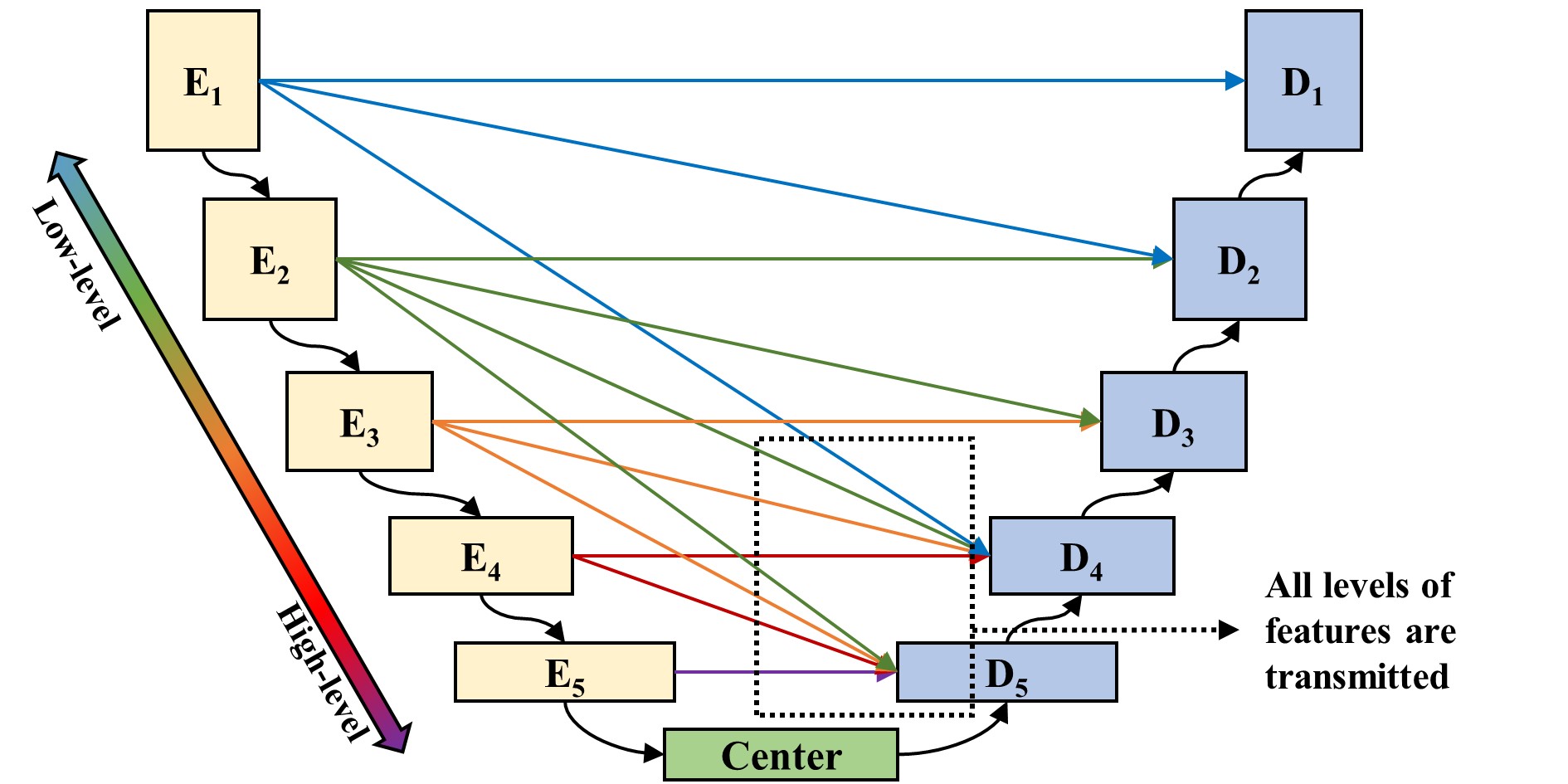}
	\caption{The concatenate layers in optimized T-Net}
	\label{img:opt_concat}
\end{figure}

In optimzized T-Net, concatenate layers are connected immediately after up-sampling layers of \textit{U}\textsubscript{\textit{D}\textsubscript{i}}\textsuperscript{1} and \textit{U}\textsubscript{\textit{D}\textsubscript{i}}\textsuperscript{2}. In case of \textit{U}\textsubscript{\textit{D}\textsubscript{4}}\textsuperscript{1}, the size of feature-map after the up-sampling is 64, and the layers of the encoders having such feature-map size is as follows.
\begin{equation}\label{(equ3.7)}
	64 = \textit{S}(\textit{E}\textsubscript{i}) = \{\textit{S}(\textit{P}\textsubscript{\textit{E}\textsubscript{4}}\textsuperscript{1}), \textit{S}(\textit{P}\textsubscript{\textit{E}\textsubscript{3}}\textsuperscript{2}),
	\textit{S}(\textit{P}\textsubscript{\textit{E}\textsubscript{3}}\textsuperscript{4}),
	\textit{S}(\textit{U}\textsubscript{\textit{E}\textsubscript{2}}\textsuperscript{3}),
	\textit{S}(\textit{U}\textsubscript{\textit{E}\textsubscript{2}}\textsuperscript{5})\}
\end{equation}
If there are layers of the same size in the encoder block, we concatenate the preceding layer only in order to reduce GPU memory consumption. Therefore, the layers concatenated with \textit{U}\textsubscript{\textit{D}\textsubscript{4}}\textsuperscript{1} are \textit{P}\textsubscript{\textit{E}\textsubscript{4}}\textsuperscript{1}, \textit{P}\textsubscript{\textit{E}\textsubscript{3}}\textsuperscript{2}, and \textit{U}\textsubscript{\textit{E}\textsubscript{2}}\textsuperscript{3}. Likewise, \textit{U}\textsubscript{\textit{D}\textsubscript{4}}\textsuperscript{2} is concatenated with \textit{P}\textsubscript{\textit{E}\textsubscript{3}}\textsuperscript{1}, \textit{P}\textsubscript{\textit{E}\textsubscript{2}}\textsuperscript{2}, and \textit{U}\textsubscript{\textit{E}\textsubscript{1}}\textsuperscript{3}. Figure \ref{img:opt_concat} shows the concatenate layers between encoder and decoder blocks. 

From Figure \ref{img:opt_concat}, we can observe that T-Net delivers various levels of features to the decoder. In case of U-Net, the encoder block connected to \textit{D}\textsubscript{4} is only \textit{E}\textsubscript{4}, but T-Net connects \textit{D}\textsubscript{4} with all blocks in the encoder except the \textit{E}\textsubscript{5}. Assume that the first concatenate layers list for \textit{D}\textsubscript{i} is \textit{M}\textsubscript{1}(\textit{D}\textsubscript{i}) and the second list is \textit{M}\textsubscript{2}(\textit{D}\textsubscript{i}). Than the lists of concatenate layers in optimized T-Net are as follow.
\begin{equation}\label{(equ3.8)}
    \begin{split}
    \textit{M}\textsubscript{1}(\textit{D}\textsubscript{5}) & = \{\textit{P}\textsubscript{\textit{E}\textsubscript{5}}\textsuperscript{1}, \textit{P}\textsubscript{\textit{E}\textsubscript{4}}\textsuperscript{2}, \textit{U}\textsubscript{\textit{E}\textsubscript{3}}\textsuperscript{3}\},\hspace{2mm}
	\textit{M}\textsubscript{2}(\textit{D}\textsubscript{5}) = \{\textit{P}\textsubscript{\textit{E}\textsubscript{4}}\textsuperscript{1}, \textit{P}\textsubscript{\textit{E}\textsubscript{3}}\textsuperscript{2}, \textit{U}\textsubscript{\textit{E}\textsubscript{2}}\textsuperscript{3}\} \\
	\textit{M}\textsubscript{1}(\textit{D}\textsubscript{4}) & = \{\textit{P}\textsubscript{\textit{E}\textsubscript{4}}\textsuperscript{1}, \textit{P}\textsubscript{\textit{E}\textsubscript{3}}\textsuperscript{2}, \textit{U}\textsubscript{\textit{E}\textsubscript{2}}\textsuperscript{3}\},\hspace{2mm}
	\textit{M}\textsubscript{2}(\textit{D}\textsubscript{4}) = \{\textit{P}\textsubscript{\textit{E}\textsubscript{3}}\textsuperscript{1}, \textit{P}\textsubscript{\textit{E}\textsubscript{2}}\textsuperscript{2}, \textit{U}\textsubscript{\textit{E}\textsubscript{1}}\textsuperscript{3}\} \\
    \textit{M}\textsubscript{1}(\textit{D}\textsubscript{3}) & = \{\textit{P}\textsubscript{\textit{E}\textsubscript{3}}\textsuperscript{1}, \textit{P}\textsubscript{\textit{E}\textsubscript{2}}\textsuperscript{2}, \textit{U}\textsubscript{\textit{E}\textsubscript{1}}\textsuperscript{3}\},\hspace{2mm}
	\textit{M}\textsubscript{2}(\textit{D}\textsubscript{3}) = \{\textit{P}\textsubscript{\textit{E}\textsubscript{2}}\textsuperscript{1}, \textit{P}\textsubscript{\textit{E}\textsubscript{1}}\textsuperscript{2}\} \\
	\textit{M}\textsubscript{1}(\textit{D}\textsubscript{2}) & = \{\textit{P}\textsubscript{\textit{E}\textsubscript{2}}\textsuperscript{1}, \textit{P}\textsubscript{\textit{E}\textsubscript{1}}\textsuperscript{2}\},\hspace{2mm} 
	\textit{M}\textsubscript{2}(\textit{D}\textsubscript{2}) = \{\textit{P}\textsubscript{\textit{E}\textsubscript{1}}\textsuperscript{1}\},\hspace{2mm} 
	\textit{M}\textsubscript{1}(\textit{D}\textsubscript{1}) = \{\textit{P}\textsubscript{\textit{E}\textsubscript{1}}\textsuperscript{1}\}
	\end{split}
\end{equation}
\textit{M}\textsubscript{2}(\textit{D}\textsubscript{i}) does not exist because the size of the feature-map is 1024 and there is no matching encoder layer with the same size. 

\subsubsection{Short-cut connections in optimized T-Net}
\begin{figure}[!tbh]
	\centering
	\includegraphics[width=\textwidth]{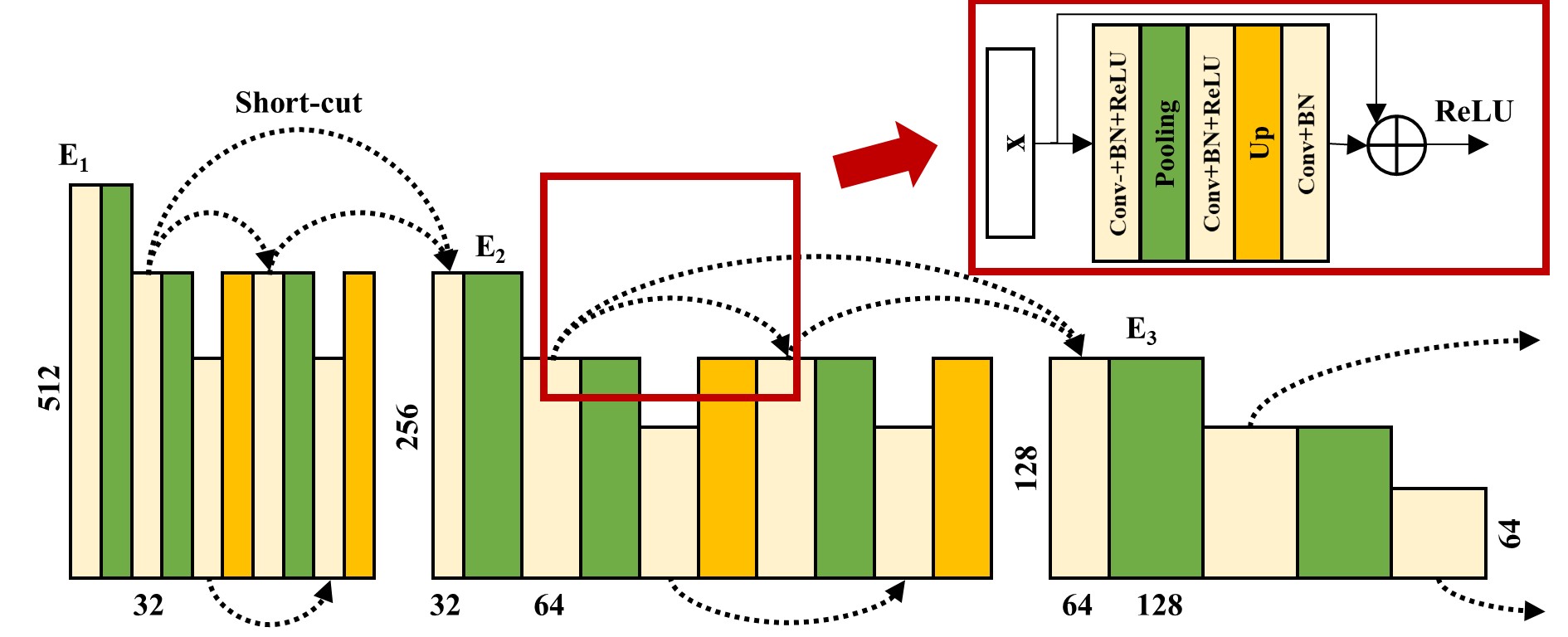}
	\caption{Short-cut connections in encoder blocks of optimize T-Net}
	\label{img:short_cut}
\end{figure}

In optimized T-Net, the number of filters in the convolutional layer of \textit{E}\textsubscript{1} starts from 32, and doubles in the next block. Short-cut connection is a concept proposed in ResNet \cite{ch23_resnet}, which has the effect of preventing the performance degradation caused by deeply stacking the convolutional layers. Therefore, recent CNN models necessarily include short-cut connections, and they also form short-cut connections that connect all the convolutional layers within a block, such as DenseNet \cite{ch23_densenet}. Short-cut connections result in the addition between layers, requiring the same number of filters as feature-maps of the same size. In case of optimized T-Net, the second and fourth, third and fifth layers of all \textit{E}\textsubscript{i} and \textit{D}\textsubscript{i} have the same number of filters as the same feature-map size, respectively. Through the equation \ref{(equ3.5)}-\ref{(equ3.7)}, the first layer in the next block has a feature-map of the same size as the second and fourth layers of the previous block. Therefore, an additional short-cut connection is possible by making only the first convolutional layer equal to the number of filters of the previous block before double the number of filters of the next block. Figure \ref{img:short_cut} shows how short-cut connections are constructed in three consecutive encoder blocks. 

Finally, the number of filters and size of feature-maps constituting each block of optimized T-Net is as follow. The number in parentheses is the size of the feature-map.
\begin{equation}\label{(equ3.9)}
    \begin{split}
	\textit{F}(\textit{E}\textsubscript{1}) & = \{32(512), 32(256), 32(128), 32(256), 32(128)\}\\ 
	\textit{F}(\textit{E}\textsubscript{2}) & = \{32(256), 64(128), 64(64), 64(128), 64(64)\} \\
	\textit{F}(\textit{E}\textsubscript{3}) & = \{64(128), 128(64), 128(32), 128(64), 128(32)\}\\ 
	\textit{F}(\textit{E}\textsubscript{4}) & = \{128(64), 256(32), 256(16), 256(32), 256(16)\} \\
	\textit{F}(\textit{E}\textsubscript{5}) & = \{256(32), 512(16), 512(8), 512(16), 512(8)\} \\
	\textit{F}(\textit{Ce}) & = \{1024(16), 512(16)\} \\
	\textit{F}(\textit{D}\textsubscript{5}) & = \{512(16), 512(32), 512(64), 512(32), 512(64)\}\\ 
	\textit{F}(\textit{D}\textsubscript{4}) & = \{256(32), 256(64), 256(128), 256(64), 256(128)\}\\ 
	\textit{F}(\textit{D}\textsubscript{3}) & = \{256(64), 128(128), 128(256), 128(128), 128(256)\} \\
	\textit{F}(\textit{D}\textsubscript{2}) & = \{128(128), 64(256), 64(512), 64(256), 64(512)\}\\ 
	\textit{F}(\textit{D}\textsubscript{1}) & = \{64(256), 32(512), 32(1024), 32(512), 32(1024)\} \\
	\textit{F}(\textit{Pr}) & = \{32(512), 4(512)\}
	\end{split}
\end{equation}
where \textit{Ce} denotes two convolutional layers in the center and \textit{Pr} refers the last two convolutional layers before final prediction.

\subsubsection{Fine-tuning T-Net for optimization}
Convolutional layers of optimized T-Net consist of convolution, batch normalization \cite{ch23_bn}, and activation function. The activation function uses softmax only in the last convolutional layer and the others use ReLU \cite{ch23_relu}. The weights of the convolution are initialized by He initialization \cite{ch3_heinit}. For up-sampling, transposed convolution \cite{ch3_transposed} and bi-linear up-sampling are the most common methods. As a result of the experiment, there was no significant difference in performance between two. The detailed performance comparison result applying each method in optimized T-Net is explained in the next section. However, we used bi-linear up-sampling because transposed convolution requires number of free parameters. 

In general, Dice Similarity Coefficient score (\textit{DSC}) is the performance metric for semantic segmentation problems. The definition of the \textit{DSC} is described in the result section. The commonly used loss function in deep learning is cross entropy loss (\textit{CELoss}). However, \textit{CELoss} has a disadvantage in that loss cannot be easily improved when a large number of background pixels are included, which corresponds to most medical image segmentation problems. Therefore, the following loss function \textit{DSCLoss} is used to provide a balance between \textit{CELoss} and \textit{DSC}. 
\begin{equation}\label{(equ3.10)}
    DSCLoss = 1 + \alpha CELoss - DSC
\end{equation}
Where $\alpha$ is a coefficient for adjusting the scale of \textit{DSCLoss} which set to 0.1 in this paper. More specifically, 1 - \textit{DSC} attempts to reduce the loss of the pixels corresponding to the vessels, and $\alpha$\textit{CELoss} tries to reduce the overall pixel loss, including the background.

In the case of the optimizer function, we used the Adam \cite{ch23_adam} and set the initial learning rate to 0.0001. In addition, we reduced the learning rate with a factor of 0.5 if the validation loss does not improve for the last 5 epochs.
%%%%%%%%%%%%%%%%%%%%%%%%%%%%%%%%%%%%%%%%%%%%%%%%%%%%%%%%%%%%%%%%%%%%%%%%%%%%%%%%%%%%%%%%%%%%%%%%%%%%%%%
\section{Results}
\label{ch3:results}
%%%%%%%%%%%%%%%%%%%%%%%%%%%%%%%%%%%%%%%%%%%%%%%%%%%%%%%%%%%%%%%%%%%%%%%%%%%%%%%%%%%%%%%%%%%%%%%%%%%%%%%
% 데이터를 어떻게 얻었고 실험 환경은 어떠한지 쓰자. (subsection을 나누지 않는다)
% ('Train+Valid: Total(', 3755, ')', {0: 1594, 1: 1036, 2: 1125})
% ('Test: Total(', 945, ')', {0: 393, 1: 271, 2: 281})
% ('Train: Total(', 3190, ')', {0: 1349, 1: 876, 2: 965})
% ('Valid: Total(', 565, ')', {0: 245, 1: 160, 2: 160})
A total of 4,700 coronary angiography images of patients who visited Asan Medical Center were evaluated. All patients participating in the study provided written informed consent and the institutional review board of Asan Medical Center approved the study. Experts with more than five years of experience split the main vessels (LAD, LCA, RCA) from the ostium to the distal site by using The CAAS QCA system (Pie Medical Imaging BV, the Netherlands) \cite{ch3_caas2}. Of the 4,700 angiography images, 1,987 (42.3\%) were LAD, 1,307 (27.8\%) were LCX, and 1,406 (29.9\%) were RCA. From the total number of 4,700 angiography images, 3,755 images (80\%) were randomly split into train-set and 945 images into (20\%) test-set with the similar class distribution. 945 test-set images consisted of 393 LAD, 271 LCX, and 281 RCA images. Validation-set consists of 565 images which correspond to about 15\% of the train-set. As a result, 3,190 train-set images consisted of 1,349 LAD, 876 LCX, and 965 RCA images. Likewise, of the 565 validation-set images, 245 images are LAD, 160 images are LCX, and 160 images are RCA.

The software and hardware environment for the evaluation are as follows. We tested on a 64GB server with two NVIDIA Titan X GPUs and an Intel Core i7-6700K CPU. The operating system is Ubuntu 16.04, and the development of the CNN model uses Python-based machine learning libraries including Keras \cite{ch23_keras}, Scikit-learn \cite{ch23_sklearn}, and TensorFlow \cite{ch23_tensorflow}.

\subsection{Evaluation setup}
\begin{figure}[!tbh]
	\centering
	\includegraphics[width=\textwidth]{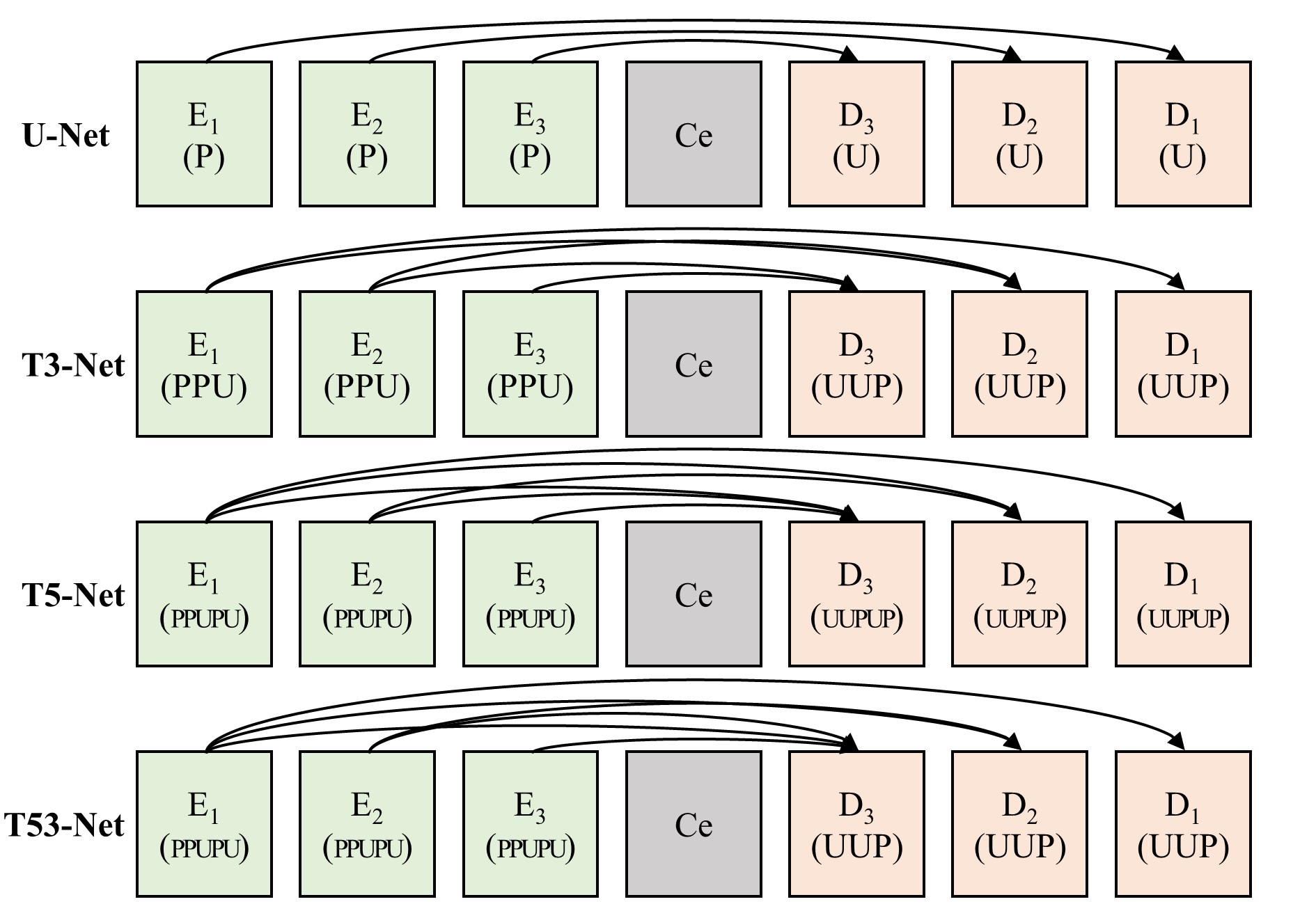}
	\caption{Schematic structures of U-Net, T3-Net, T5-Net, and T53-Net.}
	\label{img:uvst_comp}
\end{figure}

The evaluation of this paper proceeds from two perspectives. The first is to compare U-Net and T-Nets under the same conditions. In other words, it is to see how the performance of the main vessel segmentation differs in the most basic U-Net and T-Net structures. Therefore, unlike the optimized T-Net, the size of the image is resized to 256 x 256 to improve the training speed, and the depth of the encoder and decoder is fixed to 3. The learning rate was fixed at 0.0001 regardless of the validation loss, and no short-cut connection was used. Likewise, since there is no reduction in the learning rate, the max epoch is 50, which corresponds to half of the optimized T-Net. That is, the goal of the first evaluation is to compare the performance of pure U-Net with that of T-Nets, to the greatest extent, excluding other performance-enhancing factors. However, the loss function uses the same \textit{DSCLoss} as optimized T-Net because convergence is faster than using \textit{CELoss}. The T-Nets used in the comparison is T3-Net, T5-Net, and T53-Net. Unlike T-Nets, U-Net uses two consecutive convolutional layers to maintain the number of weights in each block similar to that of T3-Net. Figure \ref{img:uvst_comp} shows the schematic structure of the four models used for the U-Net and T-Nets comparisons.

The second is to evaluate the optimized T-Net that exhibits the maximum performance of the main vessel segmentation in coronary angiography. First, Opt-Net is T-Net with all the fine-tuning methods described in section \ref{ch3:optimized-t-net}. Briefly, the convolutional layer is connected with short-cut connections, the loss function is \textit{DSCLoss}, and uses bi-linear up-sampling. The T-Nets compared to Opt-Net are models with removing two methods. Opt-Net\textsuperscript{1} uses transposed convolution instead of bi-linear up-sampling and Opt-Net\textsuperscript{2} does not have the short-cut connections. Other unspecified methods apply equally to all models. More specifically, the loss function is \textit{DSCLoss}, data augmentation is applied, the learning rate has a factor of 0.5 with 5 epochs patience, and the maximum epoch is 100.

\subsection{Evaluation metrics}
The evaluation of the main vessel segmentation was based on the following three metrics: Dice Similarity Coefficient score (\textit{DSC}), sensitivity (\textit{Se}), and precision (\textit{Pr}). Dice Similarity Coefficient score, also called F1-score, means the harmonic mean of sensitivity and precision. Generally, in the semantic segmentation problem, we use the name \textit{DSC} rather than F1-score. Sensitivity, also known as the true positive rate or recall, measures the percentage of positives that are correctly identified as the main vessel. Precision measures the percentage of positives that are predicted as the main vessel. In the case of accuracy and specificity, it is not generally included in evaluation metrics for semantic segmentation problem because the class corresponding to negative is a background. These metrics are defined with the following three terminologies. However, since there are three types of main vessels, all evaluation metrics are into four different categories. That is, metrics for all types of main vessels (ALL), LAD, LCX, and RCA. 
\begin{itemize}
\setlength\itemsep{0em}
\item True Positive(\textit{TP}): The number of pixels in an angiography image correctly identified as main vessels.
\item False Positive(\textit{FP}): The number of pixels in an angiography image incorrectly identified as main vessels.
\item False Negative(\textit{FN}): The number of pixels in an angiography image incorrectly identified as background
\end{itemize}
\begin{equation}\label{(3.11)}
\begin{gathered}
	Dice\hspace{1mm}similarity\hspace{1mm}coefficient (DSC) = \frac{2\times Pr \times Se}{Pr + Se}\\
	Sensitivity (Se) = \frac{TP}{TP + FN} \times 100 (\%)\\
	Precision (Pr) = \frac{TP}{TP + FP} \times 100 (\%)
\end{gathered}
\end{equation}
In addition, loss and \textit{DSC} according to epoch are graphically represented, and visualizations of model weights and predicted mask with an actual mask are included. Detailed descriptions of the graphs and visualizations are provided in the following section.

\subsection{Evaluation results of U-Net and T-Nets}
Table \ref{tab:result_uvst} summarizes the evaluation results of U-Net and T-Nets compared under the same condition explained in the previous section.
\begin{table}[!tbh]
  \begin{center}
    \scalebox{1}{
    \begin{tabular}{|l|c|c|c|c|c|c|}
     \hline
       & \multicolumn{3}{c|}{\textit{ALL}} & \multicolumn{3}{c|}{\textit{LAD}}\\     
       Method & \textit{DSC} & \textit{Se}(\%) & \textit{Pr}(\%) & \textit{DSC} & \textit{Se}(\%) & \textit{Pr}(\%)\\
       \hline
      T53-Net & 0.807 & \textbf{81.10} & 81.89 & \textbf{0.805} & 79.96 & 82.73\\
      T5-Net & \textbf{0.815} & 80.93 & \textbf{83.74} & 0.804 & 79.77 & \textbf{82.87}\\
      T3-Net & 0.787 & 81.01 & 78.62 & 0.786 & \textbf{80.21} & 79.08\\
      U-Net & 0.720 & 75.39 & 71.52 & 0.733 & 76.25 & 72.78\\
     \hline\hline
       & \multicolumn{3}{c|}{\textit{LCX}} & \multicolumn{3}{c|}{\textit{RCA}}\\     
       Method & \textit{DSC} & \textit{Se}(\%) & \textit{Pr}(\%) & \textit{DSC} & \textit{Se}(\%) & \textit{Pr}(\%)\\
       \hline
      T53-Net & 0.730 & 75.52 & 72.68 & 0.883 & 88.08 & 89.60\\
      T5-Net & \textbf{0.753} & \textbf{75.70} & \textbf{77.08} & \textbf{0.890} & 87.58 & \textbf{91.39}\\
      T3-Net & 0.697 & 73.83 & 69.20 & 0.874 & \textbf{89.00} & 87.05\\
      U-Net & 0.565 & 63.86 & 54.10 & 0.853 & 85.27 & 86.55\\
     \hline
     \end{tabular}}
  \end{center}
  \caption{Comparison results between U-Net and T-Nets}
  \label{tab:result_uvst}
\end{table}

The highest overall \textit{DSC} was 0.815 for T5-Net, 0.08 higher than T53-Net, 0.028 higher than T3-Net and 0.095 higher than U-Net. T5-Net and T53-Net did not show significant performance differences in LAD and RCA, but T5-Net obtained 0.023 higher \textit{DSC} in LCX segmentation. Comparing T3-Net and U-Net, T3-Net was 0.067 higher than U-Net based on overall \textit{DSC}. The difference in performance between U-Net and T3-Net, which is the most similar in terms of the number of weights and structurally, demonstrates that various concatenate layers enhance performance. Considering that T3-Net has lower performance than T53-Net or T5-Net, the number of up-sampling layers of the decoder block to which the concatenate layer is connected is also important. That is, T3-Net is concatenated to the first up-sampling layer of the decoder block, but T5-Net and T53-Net are concatenated in the first and second up-sampling layers. This makes it clear that the performance of T53-Net is closer to T5, despite the fact that T53-Net is a half-mixed structure of T5 and T3.

\begin{figure}[!tbh]
	\centering
	\includegraphics[width=0.85\textwidth]{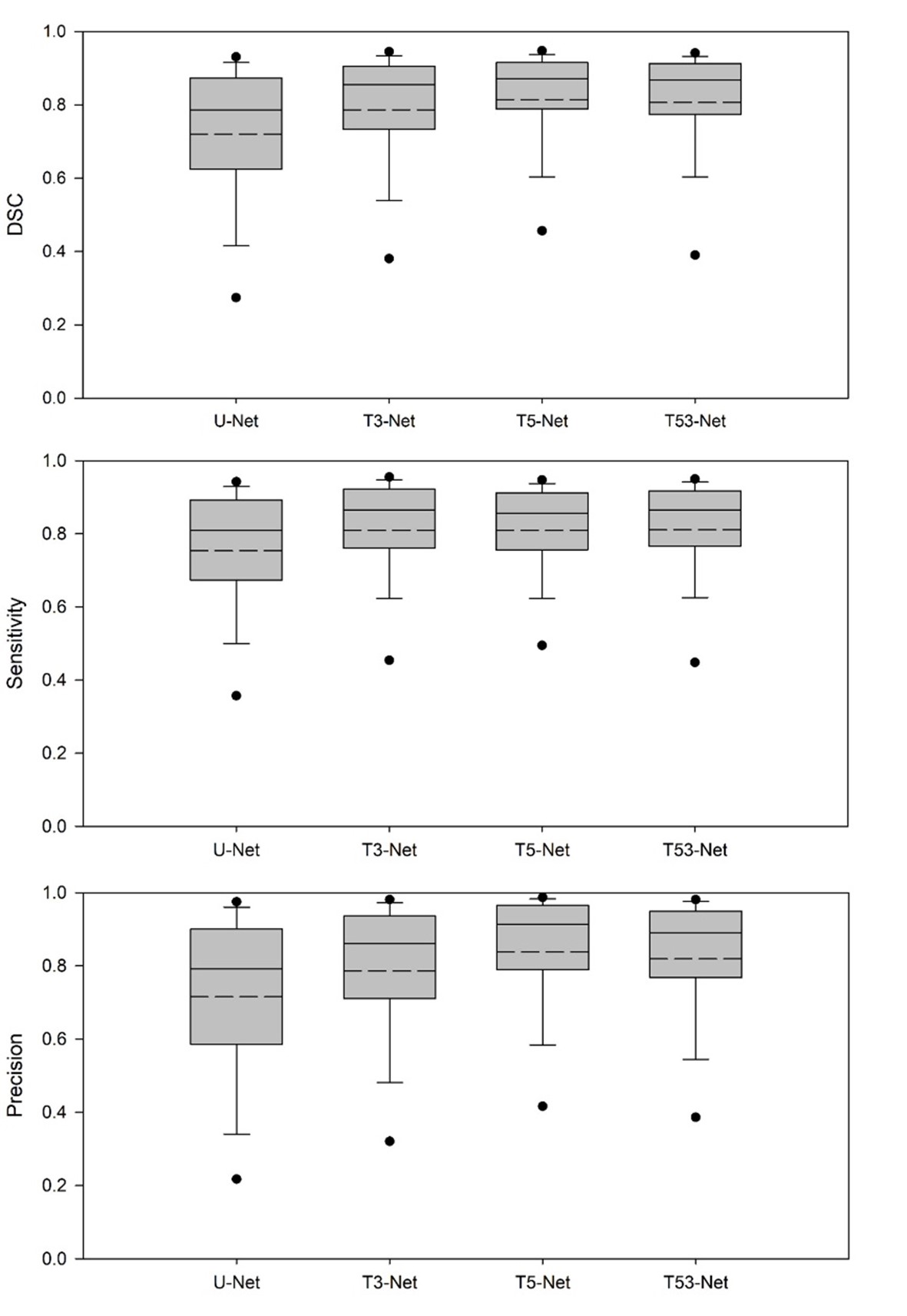}
	\caption{Box plots for comparison of U-Net and T-Nets}
	\label{img:uvst_box_dsc}
\end{figure}

The difference between U-Net and T-Net is more evident when comparing segmentation performance for each vessel. First, the smallest performance gap between T-Net and U-Net is RCA segmentation, and RCA has higher overall performance than LAD or LCX. The highest \textit{DSC} in the RCA segmentation is 0.890 in T5-Net, 0.037 higher than the lowest \textit{DSC} from U-Net. On the other hand, in LCX segmentation, T5-Net achieve 0.753 \textit{DSC} while U-Net shows only 0.565 \textit{DSC}. This means that LCX segmentation is the most difficult and RCA segmentation is the easiest problem. In other words, T-Net achieves higher performance than the U-Net in a more difficult problem, which is \textit{LCX} segmentation. Figure \ref{img:uvst_box_dsc} show box plots for overall \textit{DSC}, sensitivity, and precision for U-Net and T-Nets, respectively.

\begin{figure}[!tbh]
	\centering
	\includegraphics[width=\textwidth]{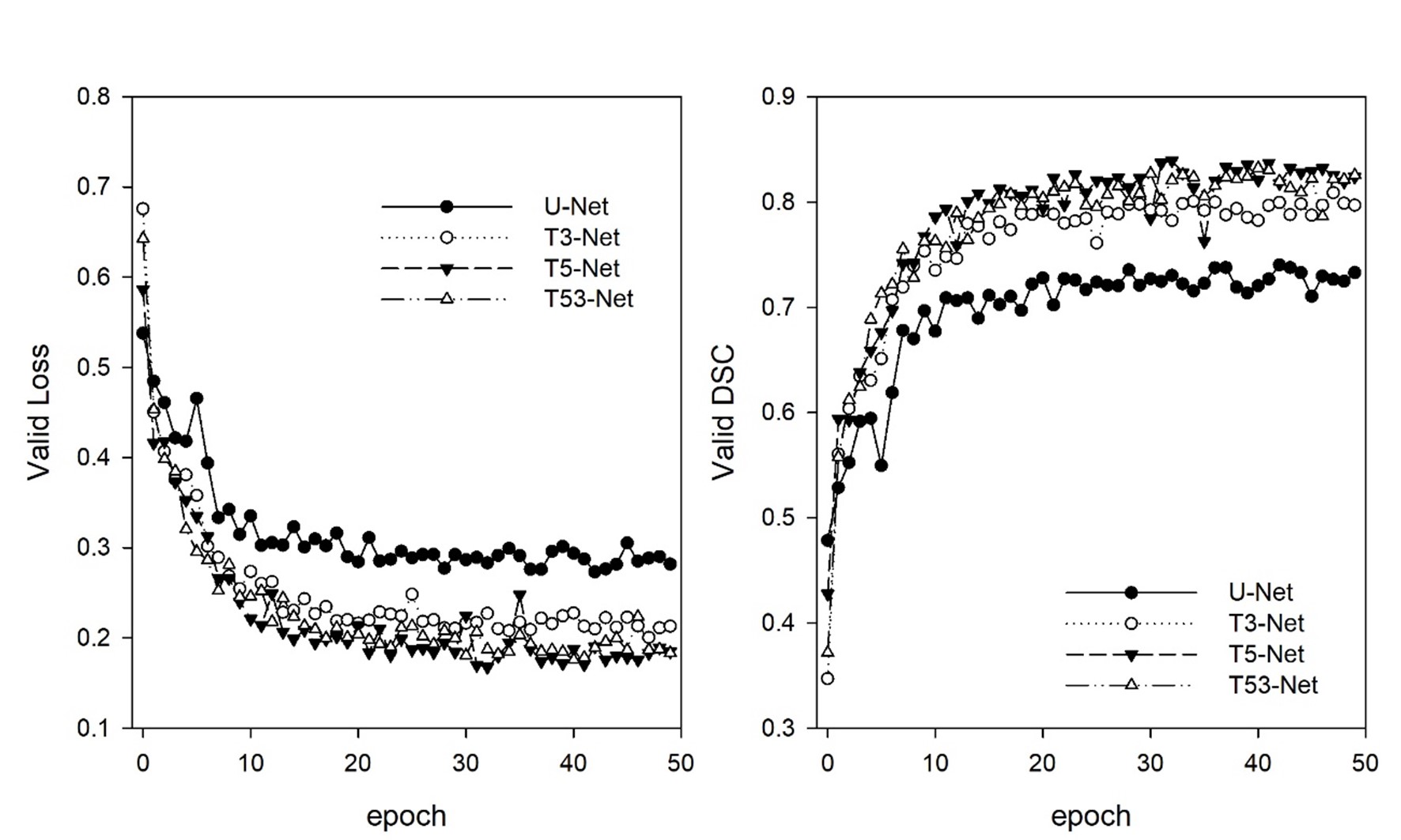}
	\caption{Validation loss and \textit{DSC} of U-Net and T-Nets}
	\label{img:uvst_valid_loss}
\end{figure}

The solid line in the box represents the median value and the dotted line represents the mean value. The rounded points above and below the box represent outliers of 5\% and 95\%, respectively. Other expressions follow the definition of the general box plot. In Table \ref{tab:result_uvst}, T53-Net and T5-Net showed similar performance on average, but the box plot shows that T5-Net achieves a fairly higher performance throughout the test-set. This is also the basis for our decision to design the optimized T-Net with T5-Net as the basis. Figure \ref{img:uvst_valid_loss} show the validation \textit{DSCLoss} and validation \textit{DSC} for each epoch while training U-Net and T-Nets. As with the above performance analysis, we can see that T-Net performs better than U-Net. In other words, we can see that \textit{DSCLoss} of U-Net converges at a higher value than those of T-Nets.

\begin{figure}[!tbh]
	\centering
	\includegraphics[width=0.9\textwidth]{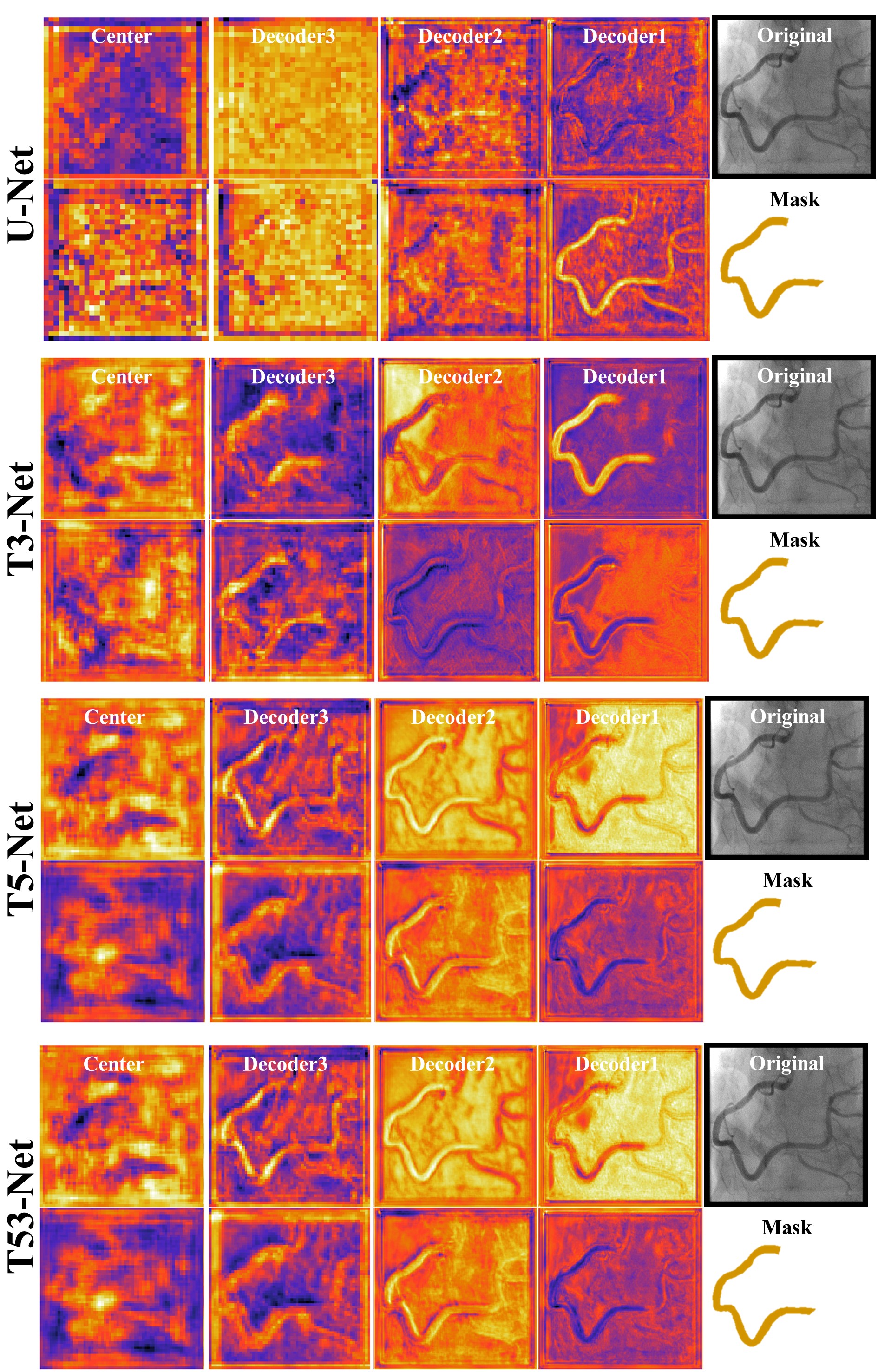}
	\caption{Visualization of weight activation for U-Net and T-Nets}
	\label{img:visualize_unet}
\end{figure}

In the comparison between U-Net and T-Nets, the last thing to evaluate is to visualize how the actual weights of each model are trained. Therefore, we visualized how the weights of the convolutional layer that constitutes each decoder block are activated for the same input image. Figure \ref{img:visualize_unet} are visualization of weight activation for U-Net, T3-Net, T5-Net, and T53-Net. 

First, we set the one of RCA vessel as the input image which shows relatively little performance difference than LAD and LCX vessels. What we want to see in Figure \ref{img:visualize_unet} is from which decoder starts to segment the mask clearly. The T-Nets begin to show the outline of the mask from the beginning of the decoder (\textit{D}\textsubscript{3}), but \textit{D}\textsubscript{3} in U-Net is almost invisible. In \textit{D}\textsubscript{2}, the weights in T-Nets are fairly clear, but those of U-Net are still unclear. And in \textit{D}\textsubscript{1}, T-Nets activate the weights similar to mask, but U-Net still activates the other vessels of the original image. This means that T-Net accurately predicts the mask from the beginning decoder block, which is possible through various concatenate layers between encoder and decoder blocks. More specifically, all the blocks of the encoder are connected to \textit{D}\textsubscript{3}, so early prediction can be performed since the low-level to high-level features extracted from the encoder are transmitted to \textit{D}\textsubscript{3}.

\subsection{Evaluation results of optimized T-Net for main vessel segmentation}
Table \ref{tab:result_opt_net} summarizes the optimized T-Net for three types of main vessel segmentation in coronary angiography.
\begin{table}[!tbh]
  \begin{center}
    \scalebox{1.1}{
    \begin{tabular}{|l|c|c|c|c|c|c|}
     \hline
       & \multicolumn{3}{c|}{\textit{ALL}} & \multicolumn{3}{c|}{\textit{LAD}}\\     
       Method & \textit{DSC} & \textit{Se}(\%) & \textit{Pr}(\%) & \textit{DSC} & \textit{Se}(\%) & \textit{Pr}(\%)\\
       \hline
      Opt-Net & \textbf{0.890} & \textbf{88.32} & \textbf{90.50} & \textbf{0.884} & \textbf{86.57} & 91.17\\
      Opt-Net\textsuperscript{1} & 0.875 & 86.24 & 89.91 & 0.878 & 85.11 & \textbf{91.56}\\
      Opt-Net\textsuperscript{2} & 0.865 & 87.10 & 87.13 & 0.855 & 84.49 & 87.88\\
     \hline\hline
       & \multicolumn{3}{c|}{\textit{LCX}} & \multicolumn{3}{c|}{\textit{RCA}}\\     
       Method & \textit{DSC} & \textit{Se}(\%) & \textit{Pr}(\%) & \textit{DSC} & \textit{Se}(\%) & \textit{Pr}(\%)\\
       \hline
      Opt-Net & \textbf{0.860} & \textbf{86.68} & \textbf{86.23} & \textbf{0.927} & 92.50 & \textbf{93.63}\\
      Opt-Net\textsuperscript{1} & 0.831 & 83.10 & 84.53 & 0.914 & 91.10 & 92.58\\
      Opt-Net\textsuperscript{2} & 0.818 & 84.74 & 80.68 & 0.923 & \textbf{93.04} & 92.28\\
     \hline
     \end{tabular}}
  \end{center}
  \caption{Evaluation results of optimized T-Net}
  \label{tab:result_opt_net}
\end{table}

The overall performance is best for Opt-Net followed by Opt-Net\textsuperscript{1} and Opt-Net\textsuperscript{2}. Opt-Net is more efficient than Opt-Net\textsuperscript{1} because the segmentation performance is better and trained with fewer free-parameters. Comparing the total number of free-parameters, Opt-Net has 43,570,372 free-parameters and Opt-Net\textsuperscript{1} has 55,883,236, which is about 28.26\% more than Opt-Net. Based on the overall \textit{DSC}, Opt-Net is the best at 0.890 followed by Opt-Net\textsuperscript{1} and Opt-Net\textsuperscript{2}. In addition, Opt-Net is always better in all three metrics, with the exception of precision for LAD and sensitivity for RCA. Opt-Net\textsuperscript{2}, which removed short-cut connections, shows a 0.025 lower \textit{DSC} than Opt-Net. That is, adding a short-cut connection has reasonable effect on performance improvement. The \textit{DSC} of Opt-Net is 0.170 higher than the \textit{DSC} of U-Net in Table \ref{tab:result_uvst}. From Table \ref{tab:result_uvst} and \ref{tab:result_opt_net}, proposed T-Net performs better than U-Net in the main vessel segmentation problem, and the optimized T-Net shows the highest performance.

The segmentation performance of each main vessel shows the highest \textit{DSC} in RCA, followed by LAD and LCX. This is in the same order as Table \ref{tab:result_uvst}, with the highest \textit{DSC} of 0.927 for RCA segmentation, which is an excellent segmentation performance. The highest \textit{DSC} for LCX is 0.860, which is 0.295 higher than U-Net, which has the lowest \textit{DSC} in Table \ref{tab:result_uvst}. In the LAD segmentation, Opt-Net's \textit{DSC} is 0.884, an improvement of 0.151 over U-Net in Table \ref{tab:result_uvst}. 

\begin{figure}[!tbh]
	\centering
	\includegraphics[width=0.95\textwidth]{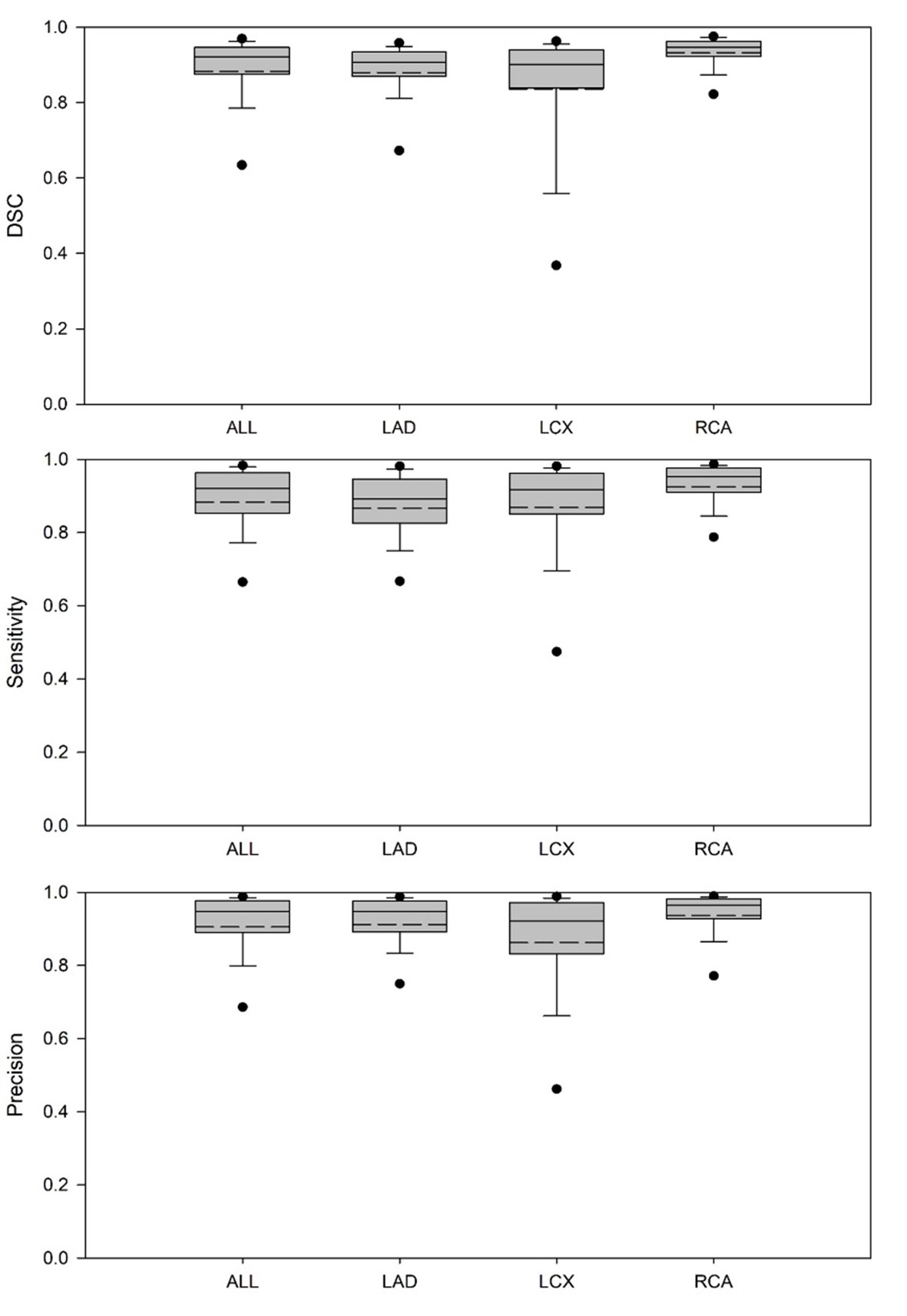}
	\caption{Box plots for optimized T-Net}
	\label{img:opt_box_dsc}
\end{figure}

Figure \ref{img:opt_box_dsc} show box plots of \textit{DSC}, sensitivity, and precision for each main vessel and overall vessel. Box plots also show that the segmentation performance of the LCX is lower than that of the other two main vessels. Considering that the number of LCX and RCA images is 1,307 and 1,406 respectively, this difference is not due to class imbalance. This is because unlike RCA, which is relatively easy to segment with no branching, LAD and LCX are more difficult to segment because they are separated from single vessel. Also, because of the small number of LCX images compared to LAD, there is performance difference between two vessels. The RCA segmentation shows high performance in the entire test-set as well as the highest average \textit{DSC} seen in the Table \ref{tab:result_opt_net}. That is, the inter-quartile range (IQR) is very narrow and is formed near 0.9 \textit{DSC}, where IQR means the difference between 75-th and 25-th percentiles. The performance of LAD segmentation is intermediate between RCA and LCX and very similar to the distribution of overall performance.

\begin{figure}[!tbh]
	\centering
	\includegraphics[width=\textwidth]{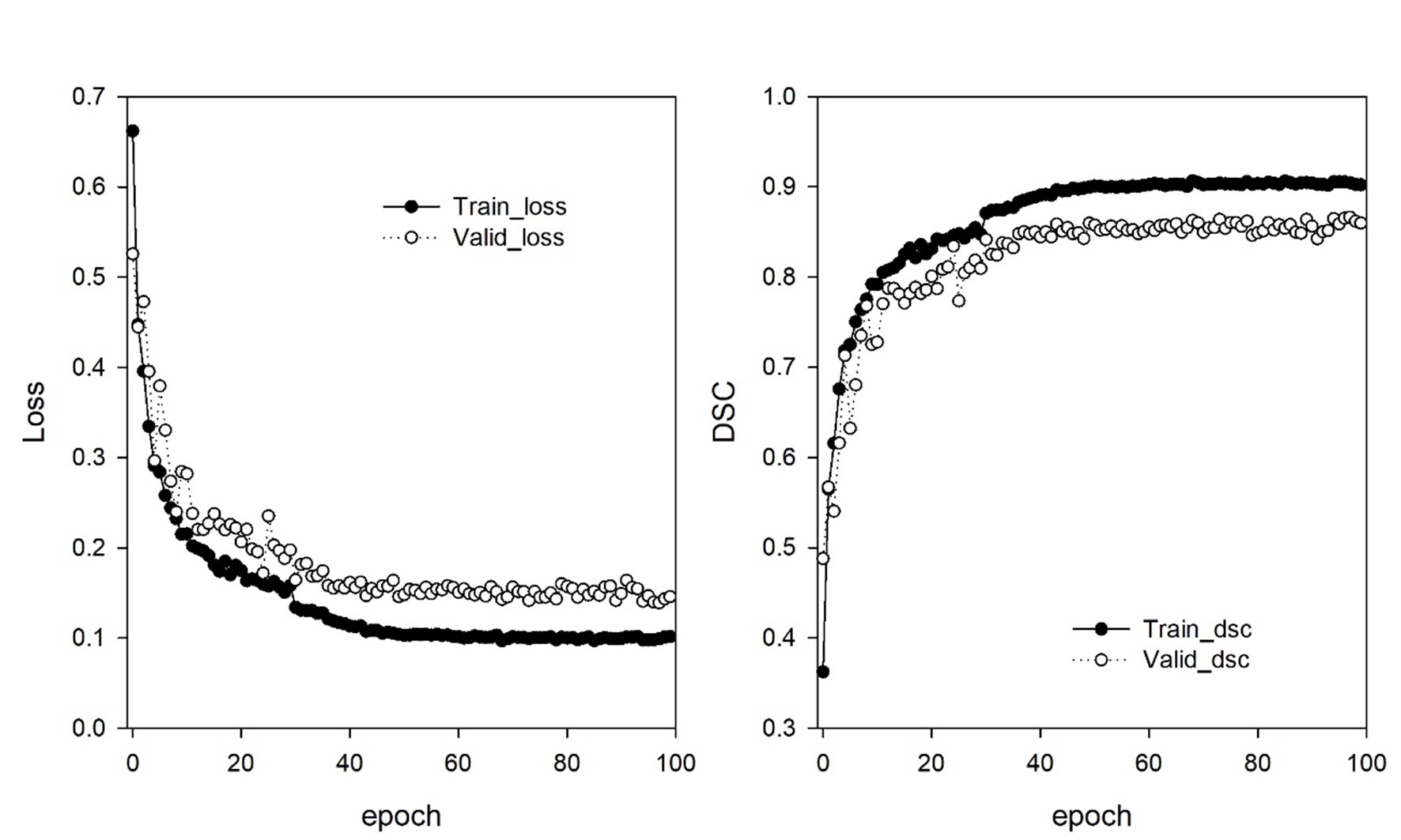}
	\caption{Loss and \textit{DSC} of optimized T-Net}
	\label{img:opt_loss}
\end{figure}

Figure \ref{img:opt_loss} show the loss and \textit{DSC} of the Opt-Net training process for each train-set and validation-set. The part where the loss and DSC fluctuate in the staircase form is the epoch where the learning rate is halved because there is no improvement in the validation \textit{DSC} during the last five epochs.

Overall, we can observe that the validation loss and \textit{DSC} improved continuously over 100 epochs. However, we have experimentally found that the minimum validation loss is formed between 80 and 100 epochs and is not improved much thereafter. Of course, if we increase the reduction factor of the learning rate and lengthen the patience epoch, the performance can be improved with longer epochs. However, we did not perform an additional evaluation with longer epochs because it took too much time because our server performance was limited. This effect is clearly observed near the 30 epoch of Figure \ref{img:opt_loss}.

Table \ref{tab:final_comp} compares the results of the proposed optimized T-Net with the previous main vessel segmentation study. Jo's study only performed LAD vessel segmentation, so no performance comparison is possible in LCX and RCA vessels. From Table \ref{tab:final_comp}, proposed T-Net in the LAD segmentation achieved 0.208 higher \textit{DSC} than Jo's approach. In addition, Jo used U-Net, which is similar to the U-Net structure used for comparison with T-Net in this paper. Under the same conditions, T5-Net showed 0.071 higher \textit{DSC} for LAD segmentation than that of U-Net. Therefore, it can be said that T-Net's segmentation performance is better than U-Net in the main vessel segmentation problem.
\begin{table}[!tbh]
  \begin{center}
    \scalebox{1}{
    \begin{tabular}{|l|c|c|c|c|c|c|}
     \hline
       & \multicolumn{3}{c|}{\textit{ALL}} & \multicolumn{3}{c|}{\textit{LAD}}\\     
       Method & \textit{DSC} & \textit{Se}(\%) & \textit{Pr}(\%) & \textit{DSC} & \textit{Se}(\%) & \textit{Pr}(\%)\\
       \hline
      Proposed & \textbf{0.890} & \textbf{88.32} & \textbf{90.50} & \textbf{0.884} & \textbf{86.57} & \textbf{91.17}\\
      T5-Net & 0.815 & 80.93 & 83.74 & 0.804 & 79.77 & 82.87\\
      U-Net & 0.720 & 75.39 & 71.52 & 0.733 & 76.25 & 72.78\\
      Jo et al \cite{ch3_vesseljo} & - & - & - & 0.676 & 60.70 & 80.00\\
     \hline\hline
       & \multicolumn{3}{c|}{\textit{LCX}} & \multicolumn{3}{c|}{\textit{RCA}}\\     
       Method & \textit{DSC} & \textit{Se}(\%) & \textit{Pr}(\%) & \textit{DSC} & \textit{Se}(\%) & \textit{Pr}(\%)\\
       \hline
      Proposed & \textbf{0.860} & \textbf{86.68} & \textbf{86.23} & \textbf{0.927} & \textbf{92.50} & \textbf{93.63}\\
      T5-Net & 0.753 & 75.70 & 77.08 & 0.890 & 87.58 & 91.39\\
      U-Net & 0.565 & 63.86 & 54.10 & 0.853 & 85.27 & 86.55\\
      Jo et al \cite{ch3_vesseljo} & - & - & - & - & - & -\\
     \hline
     \end{tabular}}
  \end{center}
  \caption{Comparison results between proposed method with previous study}
  \label{tab:final_comp}
\end{table}

\begin{figure}[!tbh]
	\centering
	\includegraphics[width=0.9\textwidth]{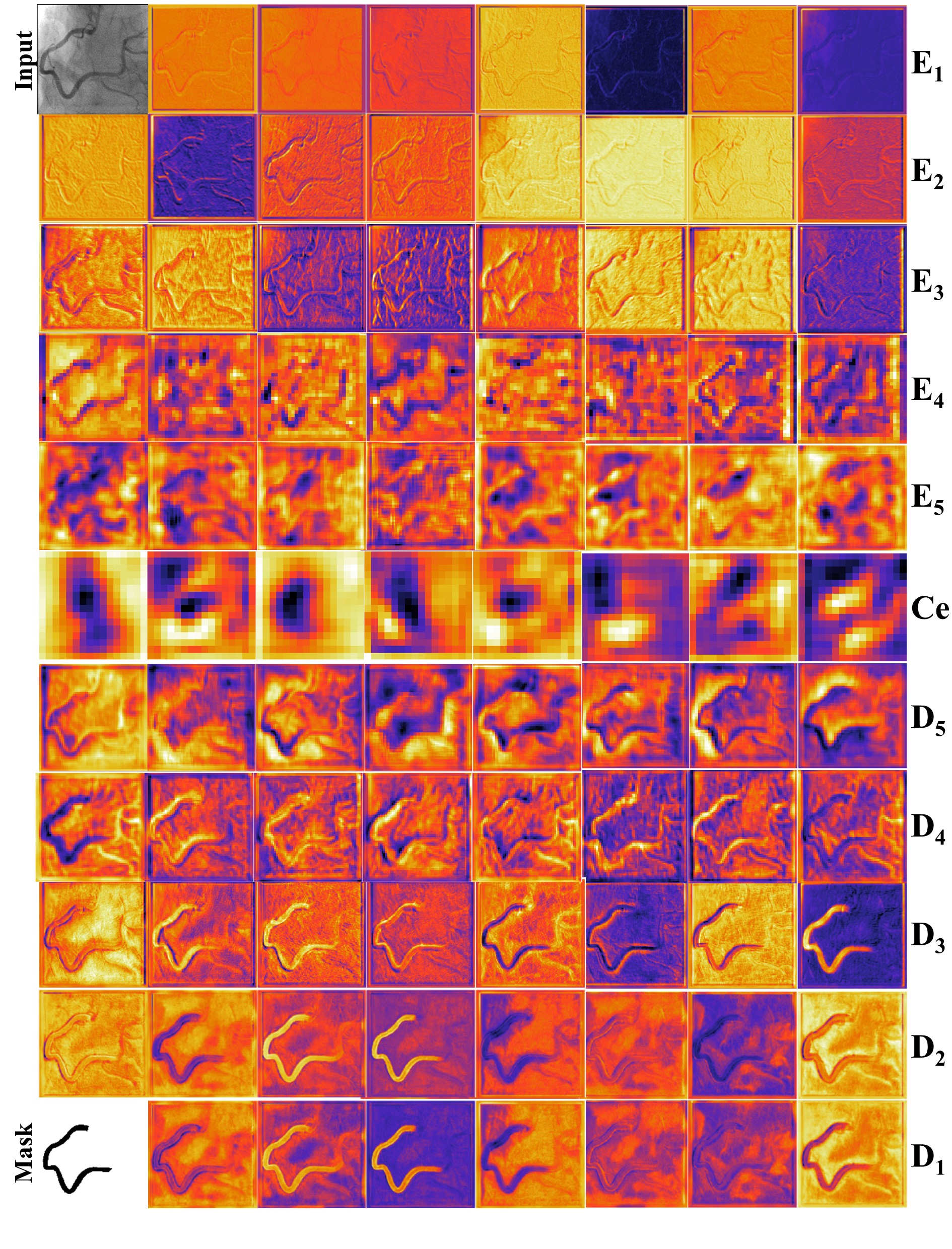}
	\caption{Visualization of weight activation for Opt-Net}
	\label{img:opt_visualize}
\end{figure}

Figure \ref{img:opt_visualize} is the visualization of weight activation for Opt-Net. Unlike Figure \ref{img:visualize_unet} with U-Net and T-Nets, we included entire blocks including encoder and decoder. We can see the process of predicting the final mask from the first encoder block through the intermediate convolution layer and then through the decoder block. In the first encoder block, low-level features such as the contour of the blood vessel are observed, and the higher-level features other than the outline of the blood vessel are extracted as the next encoder moves. When we reach the center, activation of weights is no longer similar to the original image's shape. In case of U-Net, only the information of the last encoder is concatenated to the first decoder block \textit{D}\textsubscript{5}. However, since all encoder blocks, except \textit{E}\textsubscript{1}, are connected with \textit{D}\textsubscript{5}, T-Net outputs activation close to the mask image from the beginning of restoring process. As a result, subsequent decoder blocks focus on more sophisticated restoration of the mask, resulting in higher performance. In other words, the closer the latter stage of the decoder, the clearer the differences in the activation of main vessels and background.

\begin{figure}[!tbh]
	\centering
	\includegraphics[width=0.9\textwidth]{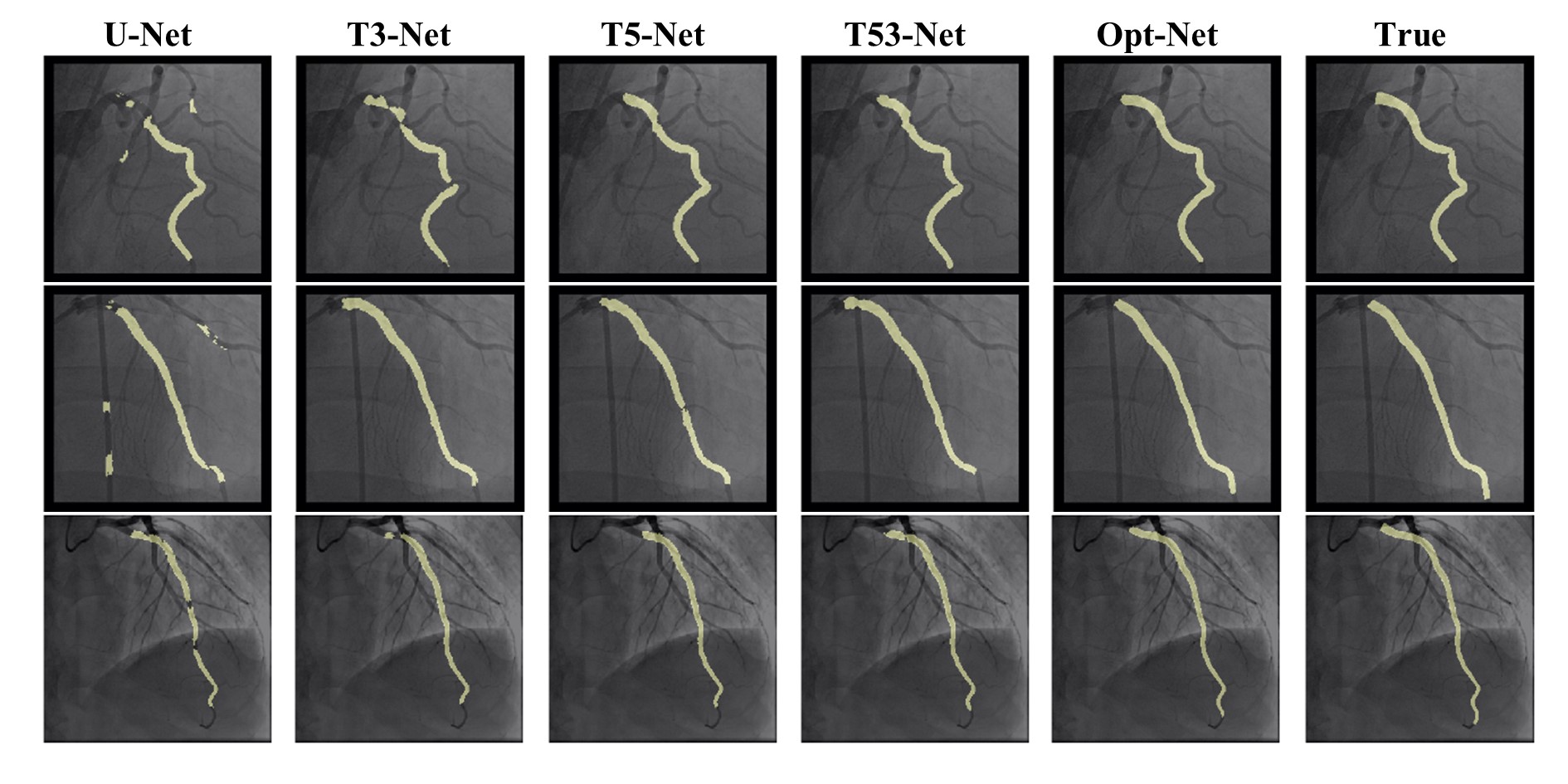}
	\caption{Visualized LAD segmentation prediction results}
	\label{img:pred_lad}
\end{figure}

\begin{figure}[!tbh]
	\centering
	\includegraphics[width=0.9\textwidth]{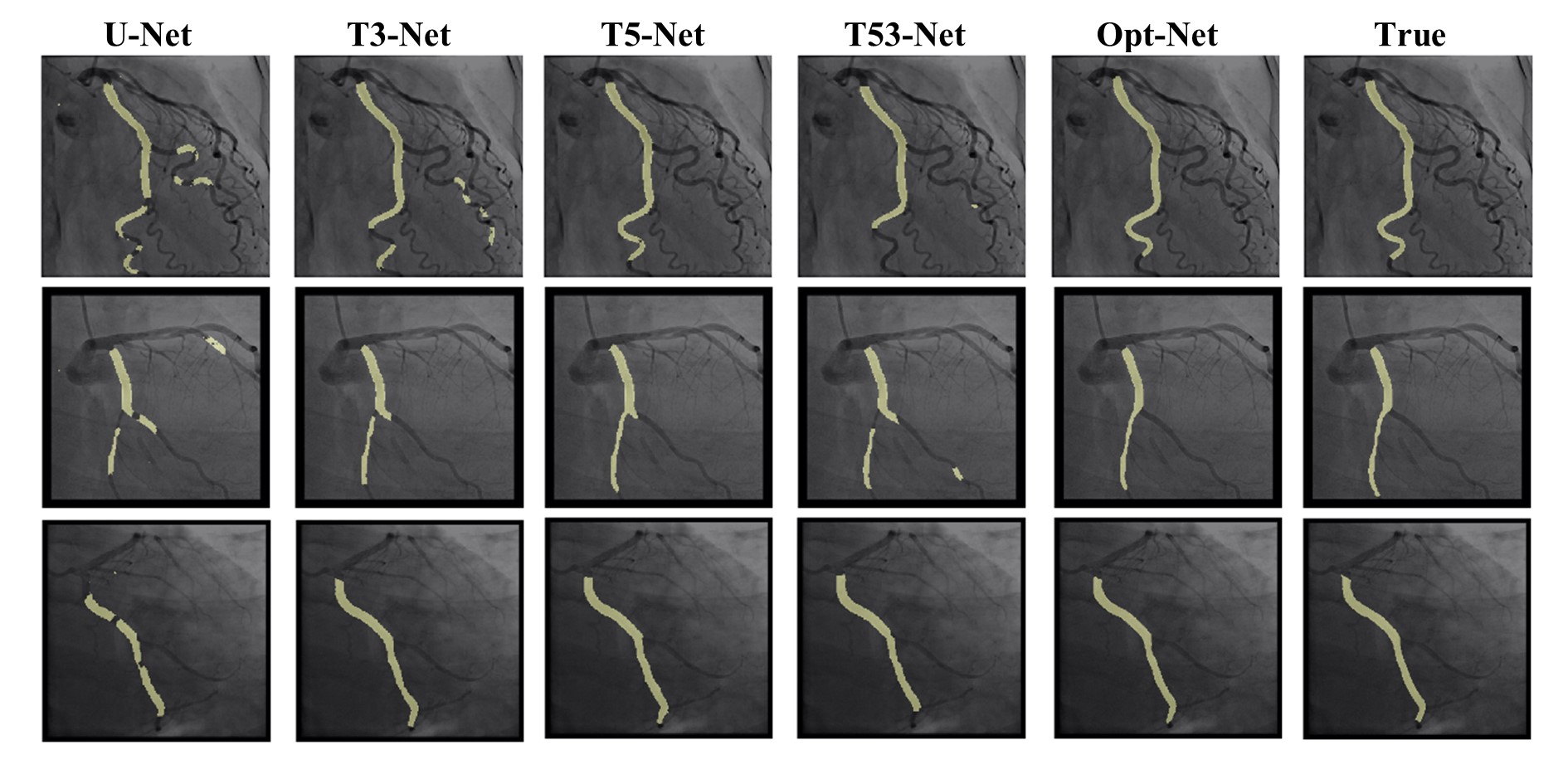}
	\caption{Visualized LCX segmentation prediction results}
	\label{img:pred_lcx}
\end{figure}

\begin{figure}[!tbh]
	\centering
	\includegraphics[width=0.9\textwidth]{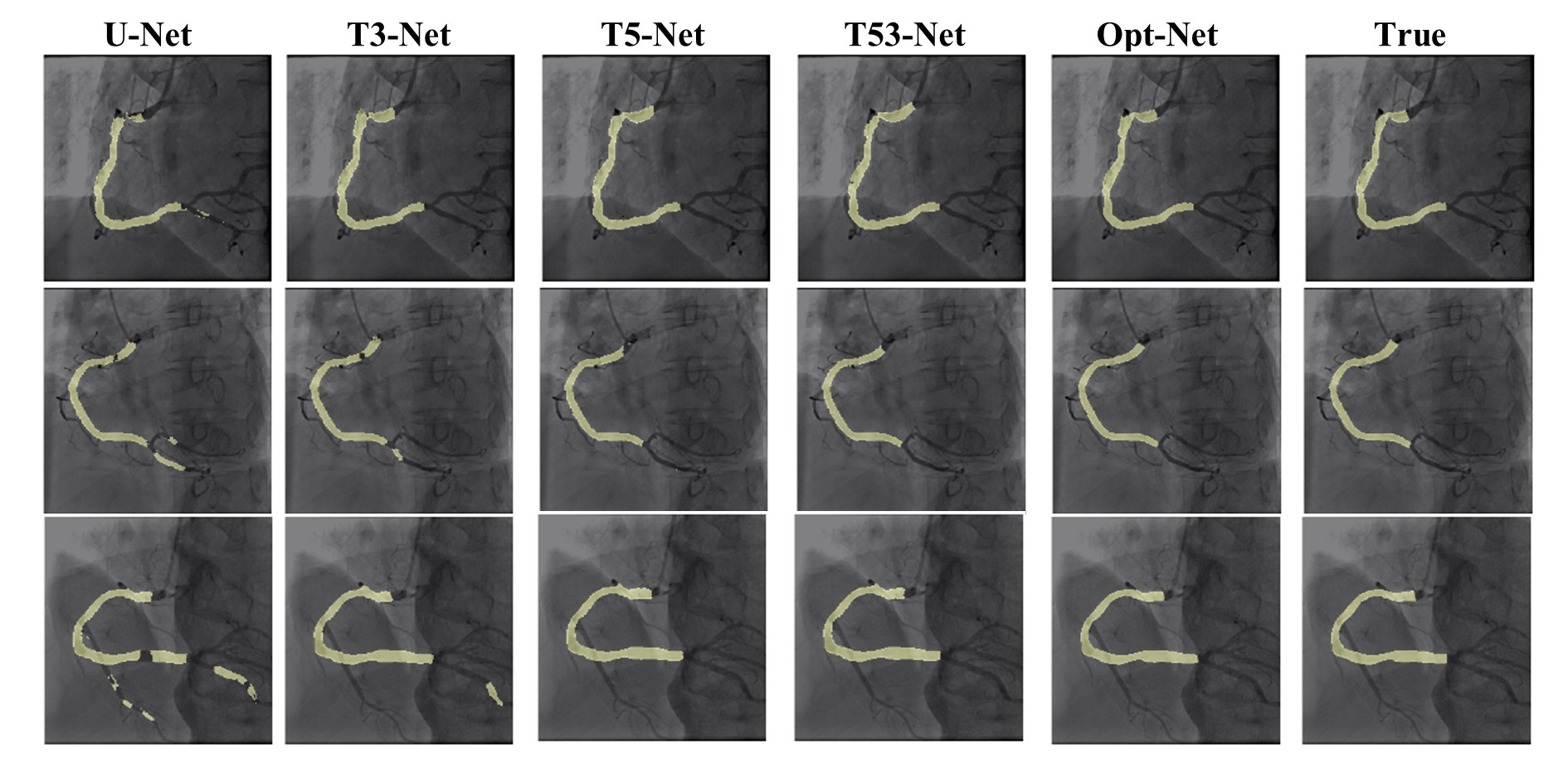}
	\caption{Visualized RCA segmentation prediction results}
	\label{img:pred_rca}
\end{figure}

Finally, we show the predicted results of LAD, LCX, and RCA segmentation of U-Net, T-Nets, and Opt-Net with actual masks in Figure \ref{img:pred_lad} through \ref{img:pred_rca}. Although these examples are not the entire test-set, but are sufficient to show the performance differences between U-Net, T-Nets, and Opt-Net. From the figures, U-Net and T-Net do not differ greatly in the role of the encoder to extract high-level feature such as main vessel position under the same condition. However, there is a performance gap in the restoration of the actual mask due to the different levels of features passed to the decoder

\section{Conclusion}
\label{ch3:conclusion}
In this paper, we proposed T-Net containing a small encoder-decoder inside the encoder-decoder structure (EDiED). T-Net overcomes the limitation that U-Net, which is the most popular model, can only have a single set of the concatenate layer between encoder and decoder block. To be more precise, the U-Net symmetrically forms the concatenate layers, so the low-level feature of the encoder is connected to the latter part of the decoder, and the high-level feature is connected to the beginning of the decoder. T-Net arranges the pooling and up-sampling appropriately during the encoder process, and likewise during the decoding process so that feature-maps of various sizes are obtained in a single block. As a result, all features from the low-level to the high-level extracted from the encoder are delivered from the beginning of the decoder to predict a more accurate mask.

We evaluated T-Net for the problem of segmenting three main vessels (LAD, LCX, RCA) in coronary angiography images. The experiment consisted of a comparison of U-Net and T-Nets under the same conditions, and an optimized T-Net for the main vessel segmentation. As a result, under the same conditions, T-Net recorded a \textit{DSC} of 0.815, 0.095 higher than that of U-Net, and the optimized T-Net recorded a \textit{DSC} of 0.890 which was 0.170 higher than that of U-Net. In addition, we visualized the weight activation of the convolutional layer of T-Net and U-Net to show that T-Net actually predicts the mask from earlier decoders. Therefore, we expect that T-Net can be effectively applied to other similar medical image segmentation problems.

Although this paper only introduces a 2-dimensional T-Net structure, the structure of 3-D T-Net is the same. Only the convolutional layer and the pooling layer change from the existing 2-D to 3-D. Therefore, We will apply 3-D T-Net to 3-D medical image segmentation problems in future work.

\section*{Acknowledgments}
This research was supported by International Research \& Development Program of the National Research Foundation of Korea(NRF) funded by the Ministry of Science, ICT\&Future Planning of Korea(2016K1A3A7A03952054) and supported by Basic Science Research Program through the National Research Foundation of Korea(NRF) funded by the Ministry of Education (2016R1D1A1A 02937565) and the National Research Foundation of Korea(NRF) grant funded by the Korea government(MSIT)( NRF-2017R1A2B3009800) and  Institute for Information \& communications Technology Promotion(IITP) grant funded by the Korea government(MSIT) (2018-0-00861, Intelligent SW Technology Development for Medical Data Analysis)

\bibliography{mybibfile}

\end{document}